\documentclass[journal]{IEEEtran}
\IEEEoverridecommandlockouts
\usepackage{amsmath}
\usepackage{soul}
\usepackage[section]{placeins}
\usepackage{verbatim} %%{\begin{comment}\end{comment}}
\usepackage{pifont}
\usepackage{times,amsmath,color,amssymb,graphicx, epstopdf, epsfig,cite,psfrag,subfigure,algorithm}
\usepackage{amsmath}
\usepackage[noend]{algpseudocode}
\usepackage{enumerate}
\usepackage{xcolor}
\usepackage{amsthm}

\newtheorem{definition}{Definition}

\newtheorem{lemma}{\underline{Lemma}}

\usepackage{color}
\begin{document}
	
	%\def\horzbar{\text{magic}}
	
	\title{
\huge Resilience of Mega-Satellite Constellations: How Node Failures Impact Inter-Satellite Networking Over Time?


\thanks{This work is supported by the National Natural Science Foundation of China under Grant 62171456, 
	in part by NSF  ECCS-2302469, CMMI-2222810, Toyota. Amazon and Japan Science and Technology Agency (JST) Adopting Sustainable Partnerships for Innovative Research Ecosystem (ASPIRE) JPMJAP2326, 
	in part by Singapore Ministry of Education (MOE) Tier 1 (RG87/22 and RG24/24), the NTU Centre for Computational Technologies in Finance (NTU-CCTF), and the RIE2025 Industry Alignment Fund - Industry Collaboration Projects (IAF-ICP) (Award I2301E0026), administered by A*STAR, 
and in part by the Fund of China Scholarship Council. 
%(\textit{Corresponding author:  Zhou~Zhang}.)
}

\thanks{B.~Guo is with the School of Telecommunication Engineering, Xidian University, Xi'an 710071, China, Tianjin Artificial Intelligence Innovation Center (TAIIC), Tianjin 300401, China, and also with the Pillar of Information Systems Technology and Design, Singapore University of Technology and Design, Singapore 487372 (e-mail: bqguo@stu.xidian.edu.cn). }

\thanks{Z.~Xiong is with the School of Electronics, Electrical Engineering and Computer Science, Queen's University Belfast, BT9 5BN Belfast, U.K. (e-mail: z.xiong@qub.ac.uk).}

\thanks{Z.~Zhang is with College of Computer Science and Electronic Engineering, Hunan University, Changsha 410012, China (e-mail: zt.sy1986@163.com).}

\thanks{B.~Li  is with the School of Mathematics and Statistics, Xidian University, Xi' an 710071, China (e-mail: bs.li@stu.xidian.edu.cn)}

\thanks{D.~Niyato is with the College of Computing and Data Science, Nanyang Technological University, 639798, Singapore (e-mail: dniyato@ntu.edu.sg).}

\thanks{C.~Yuen is with the School of Electrical and Electronics Engineering, Nanyang Technological University, Singapore 639798 (e-mail: chau.yuen@ntu.edu.sg).}

\thanks{Z.~Han is with the Department of Electrical and Computer Engineering at the University of Houston, Houston, TX 77004 USA, and also with the Department of Computer Science and Engineering, Kyung Hee University, Seoul, South Korea, 446-701 (e-mail: hanzhu22@gmail.com).}

 \author{Binquan~Guo, Zehui~Xiong,~\IEEEmembership{Senior Member, IEEE}, Zhou~Zhang,~\IEEEmembership{Member, IEEE}, Baosheng~Li,\\Dusit~Niyato,~\IEEEmembership{Fellow,~IEEE}, Chau~Yuen,~\IEEEmembership{Fellow,~IEEE}, and Zhu~Han,~\IEEEmembership{Fellow,~IEEE}}
}

	\maketitle
	
	% \textcolor[rgb]{1,0,0}{NOTE: .}
	
	\begin{abstract}
		
%	You advertise your benefits. not features

Mega-satellite constellations have the potential to leverage inter-satellite links to deliver low-latency end-to-end communication services globally, thereby extending connectivity  to underserved regions.  However, harsh space environments make satellites vulnerable to failures, leading to node removals that disrupt inter-satellite networking. With the high risk of satellite node failures, understanding their impact on end-to-end services is essential. This study investigates the importance of individual nodes on inter-satellite networking and the resilience of mega satellite constellations against node failures. We represent the mega-satellite constellation as discrete temporal graphs and model node failure events accordingly. To quantify node importance for targeted services over time, we propose a service-aware temporal betweenness metric. Leveraging this metric, we develop an analytical framework to identify critical nodes and assess the impact of node failures. The framework takes node failure events as input and efficiently evaluates their impacts across current and subsequent time windows. Simulations on the Starlink constellation setting reveal that satellite networks inherently exhibit resilience to node failures, as their dynamic topology partially restore connectivity and mitigate the long-term impact. Furthermore, we find that the integration of rerouting mechanisms is crucial for unleashing the full resilience potential to ensure rapid recovery of inter-satellite networking.

\vspace{-3 mm}

	\end{abstract}


 \vspace{-3 mm}

\section{Introduction}  \label{sec:intro}

\vspace{-3 mm}
\subsection{Motivation}

Recent advancements in space technology, particularly the advent of reusable rockets, have dramatically lowered satellite deployment costs, leading to the rapid expansion of low Earth orbit (LEO) communication satellite constellations \cite{le2024survey, di2019ultra, al2022next}. These systems are aimed at expanding wireless internet connectivity to remote and underserved regions, with the potential to reach up to $70\%$ of the Earth's surface that currently remains unconnected \cite{hassan2020dense}. 
Notably, SpaceX \cite{SPACEFLIGHT-NOW2022} stands out as a pioneering force in this arena, developing the expansive Starlink constellation with tens of thousands of satellites designed to provide seamless global wireless communication. 
Initially, Starlink employs a ``bent-pipe'' architecture that relies on ground stations for data relay \cite{zhang2023time}. While research suggests that using ground station-based relays in satellite constellations may offer latency advantages over fiber optics \cite{handley2019using}, the deployment of ground stations worldwide is fraught with challenges. In particular, establishing them in remote areas such as open oceans, deserts, and foreign territories involves considerable logistical and financial obstacles, which are further compounded by varying legal restrictions in different regions \cite{guo2024time}. To overcome these limitations, advanced inter-satellite link (ISL) technologies, including laser communications \cite{zech2015lct}, are currently being deployed to enable direct satellite-to-satellite wireless communication and support multi-hop inter-satellite routing, thereby reducing reliance on ground infrastructure \cite{guo2023online, cao2023edge, guo2024lightweight}. If fully  implemented, ISLs can facilitate low-latency inter-satellite networking, potentially further decreasing end-to-end propagation delays to the order of  tens of milliseconds and paving the way for globally accessible wireless communication services in the near future \cite{lai2022spacertc, guo2024enhanced, dakic2023delay, huang2024fair}.

However, in the harsh and complex space environment, satellite nodes are inherently prone to failures \cite{zhang2023modeling, yue2023low}. Indeed, LEO satellites are exposed to high radiation levels, extreme temperature fluctuations, solar wind, and frequent micro-collisions with space debris, all of which contribute to hardware degradation and increase the risk of software malfunctions  \cite{yue2020outage, li2024performance}. 
Unlike ground-based wireless infrastructure, damaged or failing satellites cannot be easily repaired, making any failure event effectively irreversible and limiting response options for network operators. According to publicly disclosed statistics, as of the end of October 2024, Starlink has launched a total of $7,073$ satellites, with $1,788$ currently inactive and $604$ burned, indicating a node failure rate of approximately $25.3\%$. In other words, for every $4$ satellites launched, approximately $1$ satellite is either inactive or in a malfunction state. 
As the scale of satellite constellations grows and satellites age, the risk of node failures will rise significantly.  This poses a substantial challenge for sustaining reliable wireless communication services, particularly when catering to a vast number of ground users. Therefore, understanding how node failures impact inter-satellite networking is essential for effective network design, maintenance, and evaluation. 
Moreover, satellite failures can lead to substantial economic implications due to both direct losses and service disruptions. Direct losses include the high costs associated with satellite manufacturing and launch, which can reach millions of dollars per satellite, as well as expenses for premature replacements or network reconfigurations. Service disruptions may result in user attrition, penalties for breaching service-level agreements, and significant revenue losses, particularly in sectors that rely heavily on service reliability, such as aviation, maritime operations, and government communications. Consequently, satellite failures inevitably increase the operational burdens for network operators and may further escalate costs due to the potential degradation in networking performance for users.

Intuitively, for a given end-to-end communication service between ground cells, the failure of any node along the service path will lead to a link disruption, interrupting data transmission and necessitating rerouting to an alternative path. However, due to the dynamic and periodical movement of satellite nodes, both the importance of nodes and impact of node failures change over time. Additionally, the mesh-like structure of LEO satellite constellations causes satellites to be periodically replaced by adjacent satellites in the same orbit, making the analysis of node failure impacts distinct from that in traditional static networks. Specifically, we identify that the impact of node failures on inter-satellite networking has two key features that complicate analysis:
\begin{itemize}
  
  \item 
 \textbf{Time-dependent: }  Due to the continuous movement of satellites, the importance of a satellite node fluctuates over time. A node that is critical for inter-satellite networking at one time may become less significant as satellites shift positions. Consequently, the impact of a node failure on inter-satellite networking is often confined to specific time intervals, reflecting the temporal nature of satellite connectivity.

  \item 
 \textbf{Service-dependent:} 
Within a given time period, a non-uniform distribution of service demands across the Earth leads to some satellites actively engaged in operations while others remain idle. Consequently, satellite nodes are not consistently active for all services at all times. Therefore, the failure of a specific node will disrupt only those services dependent on it, leaving others unaffected. This necessitates a refined analytical approach to assess the service-dependent impact of each node failure, identifying which specific services are affected based on the failed node's operational role and engagement.

\end{itemize}

% \vspace{-3 mm}
\begin{figure}[t!]
	\centering
 	\includegraphics[width=88mm]{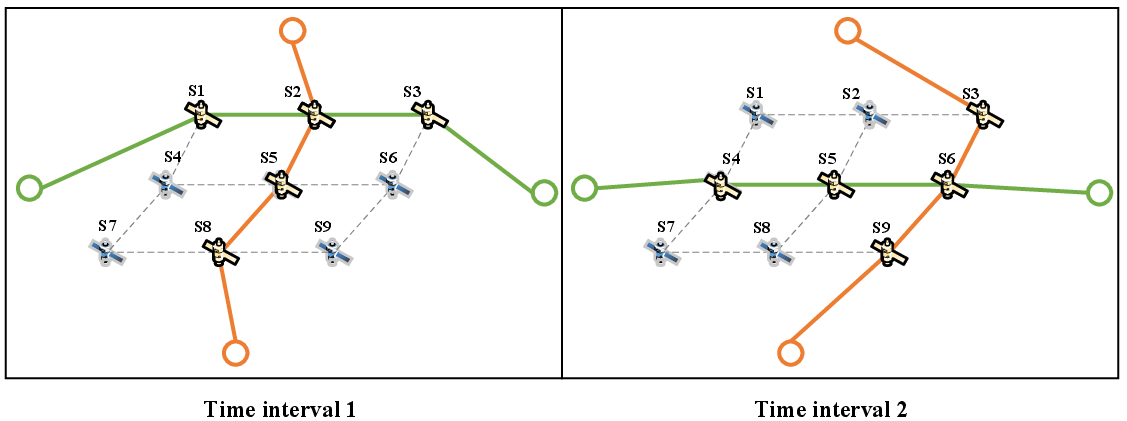}\\
% 	 \vspace{-3 mm}
	\caption{An example illustrating the time-dependent and service-dependent features of node failure impact in an  mega-satellite constellation, with two end-to-end services across two time intervals. Circles represent ground cells, and highlighted satellites are actively engaged in services. }
\label{fig:two_factors_of_node_failure_impact}
\end{figure}

Figure \ref{fig:two_factors_of_node_failure_impact} illustrates the two key features of node failure impact using a simple example, involving two end-to-end services across two time intervals. In this example, green circles represent a pair of ground cells in \textit{Service 1}, while orange circles indicate those in \textit{Service 2}. Highlighted satellites represent active satellites along the service path for each service.
In the first time interval, satellites \textit{S1}, \textit{S2} and \textit{S3} form the service path for \textit{Service 1}, while satellites \textit{S2}, \textit{S5}, and \textit{S8} form the service path for \textit{Service 2}. In the second time interval, the nodes forming the service path for \textit{Service 1} change to \textit{S4}, \textit{S5} and \textit{S6}, while those for \textit{Service 2} shift to \textit{S3}, \textit{S6} and \textit{S9}. Obviously, if satellite \textit{S1} fails, its disruption is confined to the first time interval, as satellite \textit{S4} takes over its role in the second time interval. However, the failure of satellite \textit{S3} will disrupt both the \textit{Service 1} and \textit{Service 2},  in the first and second time intervals, respectively. Conversely, a failure of satellite \textit{S7} has no effect on either service. This example highlight  the node failure impacts on inter-satellite networking depends on both the time and the services.

Given the above-mentioned two features, an analytical model is needed to accurately capture both the temporal and service-related characteristics of node failure impact. Existing studies focuses either on preventing node failures or on designing resilient methods to mitigate the effect of failures on services. However, there remains a gap in understanding how satellite node failures impact inter-satellite networking. In this work, we aim to understand the impact of satellite node failures by addressing the following key questions:
\begin{itemize}
  
  \item  \textbf{Q1:} How does the importance of a satellite node to end-to-end communication services change over time?

  \item  \textbf{Q2:}  How do node failures impact end-to-end services, and how well can satellite networks recover from them?
  
\end{itemize}

By exploring both questions, our work aims to provide deeper insights into the importance of nodes and the impact of node failures in dynamic satellite networks over time. It also evaluates the robustness of mega-satellite constellations against failures, informing resilience and recovery strategies.

\vspace{-3 mm}
\subsection{Related Work}

A substantial body of research has focused on leveraging inter-satellite networking within mega-satellite constellations to enhance data transmission capabilities, such as \cite{tao2023joint, chen2022leo,chen2024shortest, handley2018delay, chen2021analysis}. However, these studies do not explore scenarios involving node failures in inter-satellite networking. Notably, the authors in \cite{chen2021analysis} proposed a theoretical model to assess the hop count of inter-satellite paths between ground users, providing preliminary insights into topology and routing design. Their follow-up studies \cite{chen2022leo} and \cite{chen2024shortest} further investigated shortest path-based inter-satellite networking within satellite constellations without considering node failures.
Meanwhile, the authors in \cite{handley2018delay} utilized a Starlink-based simulator to verify that inter-satellite networking can achieve lower latency compared to terrestrial networks for distances exceeding 3,000 km. It highlights the inherent resilience of mega-satellite networks but does not quantitatively examine the implications of node failures on inter-satellite networking. As a result, these studies  have not delved into the risks associated with node failures, leaving a gap in understanding their impact on inter-satellite networking.

On the other hand, there are also many works that study the robustness and resilience of satellite networks. However, these studies have either focused on reducing internal and external risks, or on developing resilient networking strategies to mitigate the impact of node failures. For instance, the authors in \cite{wang2019pkn} presented methods to enhance the survivability of LEO satellite networks by protecting critical nodes. The authors in \cite{han2021secure} proposed a secure architecture designed to safeguard satellite systems against jamming attacks. These works enhance LEO satellite reliability in various aspects, and their contributions are summarized in \cite{yue2023low}. In comparison, numerous studies propose strategies to improve the resilience of satellite networks when nodes fail, such as \cite{li2022secure}, \cite{lai2023achieving}, and \cite{zhang2024resilient}.
In particular, \cite{lai2023achieving} introduces an adaptive hybrid routing scheme to efficiently manage predictable failures, alongside a location-guided protection strategy to swiftly address unexpected failures. The authors in \cite{zhang2024resilient} proposed a resilient shortest path first directional routing algorithm for handling multiple node failures in inclined LEO mega constellations, which guarantees $100\%$ reachability against any 
$k$ node failures. While these approaches can reduce the likelihood of failures or responding effectively post-failure, they do not examine how individual node failures affect inter-satellite networking.

With respect to analyzing the impact of node failures on networking performance, betweenness centrality is a widely used metric \cite{boccaletti2006complex, costa2007characterization, lu2020structural}. Originally developed for social network analysis, it has been adapted to identify central nodes in wireless communication networks. While effective, this metric evaluates node importance from a topological perspective based on an all-to-all service assumption, neglecting both time and service-dependent factors. To address this limitation, the authors in \cite{zhao2020efficient} introduced the efficient betweenness, which incorporates service-specific factors but ignores the evolving effects of node importance over time. Similarly, the authors in \cite{xu2023robustness} examined the robustness of satellite constellation networks using a single snapshot, disregarding the temporal changes in the effects of node failures. To incorporate time-dependent nature of node failure impact, the authors in \cite{zhang2017temporal} proposed the temporal centrality which captures the evolving importance of nodes, but does not account for service-dependent factor. However, these approaches fail to integrate both time- and service-dependent aspects of node importance.

This paper aims to bridge these gaps by proposing a service-specific temporal betweenness metric that reveals both the time-varying and service-dependent impacts of node failures in dynamic satellite constellations. This approach facilitates a more comprehensive analysis of how node failures affect wireless communication services, paving the way for the development of resilient mega-satellite constellations.

\vspace{-3mm}
\subsection{Our Contributions and Insights}

To investigate the impact of node failures on inter-satellite networking from the perspective of two identified factors, we represent mega-satellite constellations as discrete temporal graphs and model node failure events accordingly. We propose a service-aware temporal betweenness (SATB) metric, which quantifies the importance of each satellite node for designated wireless communication services. Leveraging this metric, we develop an analytical framework to identify critical nodes and evaluate the impact of node failures on inter-satellite networking across different time windows. 
Simulations on the Starlink constellation reveal that satellite networks inherently exhibit resilience to node failures, as their dynamic topology partially restore service connectivity and mitigate the long-term impact of node failures. Furthermore, we find that the integration of rerouting mechanisms is crucial for unleashing the full resilience potential of the satellite constellation to ensure rapid recovery of inter-satellite networking. 
Our contributions, insights, and limitations are summarized as follows:

\begin{itemize}
\item \textbf{Temporal Graph-Based Modeling of Satellite Node Failures}. We are the first to model the risks of node failures using discrete temporal graphs. We analyze node failure events from a networking perspective and reveal that their impacts on inter-satellite networking are both time-dependent and service-dependent. Unlike existing works focused on preventing or mitigating failures, our model acknowledges the inevitability of node failures and captures their evolving impact across time windows and services, providing a more comprehensive understanding of the risks posed by node failures in satellite networks.

\item \textbf{Time-Varying Node Importance Metric and Analytical Framework}. We propose the SATB metric to capture time-varying node importance in satellite constellations for designated services. This metric enables the analytical framework that quantitatively evaluates the impact of node failures on inter-satellite networking. 
By modeling connectivity, path delays, and affected nodes, the framework quantifies the evolving impact of node failures and identifies vulnerabilities and performance bottlenecks over time and across services.

\item \textbf{Real-World Starlink System Setting Evaluation and Network Design Insights}. Simulations on the Starlink system setting using the proposed framework provide two insights for satellite system design. On the one hand, we find that satellite networks exhibit inherent resilience to node failures, as their dynamic topology partially restores service connectivity and limits long-term impacts. On the other hand, our analysis demonstrates the vital role of rerouting mechanisms in large-scale failures, with their integration into the network's inherent resilience being essential for unleashing the full resilience potential of satellite constellations and enabling rapid recovery to near pre-failure performance.

\item \textbf{Limitations.} Our model considers satellite networks with end-to-end services supported by the widely adopted shortest-path routing scheme. While these assumptions reflect practical satellite network settings, they may limit applicability to other scenarios involving multi-user collaboration or alternative routing strategies.

\end{itemize}

The remainder of this paper is organized as follows. Section \ref{sec:system_model} describes the system model and problem statement. In Section \ref{sec:framework}, we propose a node failure impact analysis framework. The simulation results are presented and discussed in Section \ref{sec:evaluation}. Finally, conclusions are drawn in Section \ref{sec:conclusion}.

 \vspace{-3 mm}
\section{System Model and Problem Statement} \label{sec:system_model}

\subsection{ISL-Empowered Mega-Satellite Networking Model}

\begin{figure}[t!]
	\centering
 	\includegraphics[width=66mm]{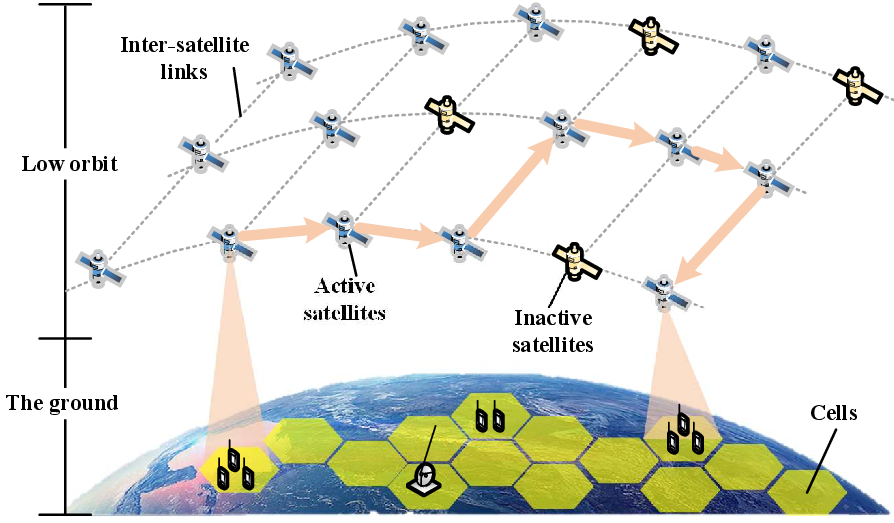}\\
% 	 \vspace{-3 mm}
	\caption{The scenario of the mega-satellite constellation with inter-satellite links (ISLs) for end-to-end routing. Due to the continuous movement of LEO satellites relative to ground cells, routing algorithms inherently require frequent updates over time. However, when some satellite nodes experience failures, the routing algorithms must dynamically avoid these inactive nodes to maintain end-to-end wireless communication. Consequently, node failures, especially those critical nodes, can significantly impact inter-satellite networking. }
\label{fig:satellite_networks_with_ISLs}
\end{figure}

\textbf{Mega-Satellite Constellation Model.} 
As is illustrated in Figure \ref{fig:satellite_networks_with_ISLs}, our system model focuses on a mega-satellite constellation with ISLs, providing end-to-end wireless communication services to ground users within globally distributed cells.
The LEO satellites are represented by the set \( \mathcal{S}= \{\mathbb{S}_1, \mathbb{S}_2, \mathbb{S}_3, \ldots, \mathbb{S}_{|\mathcal{S}|} \}\), while the Earth's surface is partitioned into multiple cells, denoted as the set \(  \mathcal{C} = \{\mathbb{C}_1, \mathbb{C}_2, \mathbb{C}_3, \ldots, \mathbb{C}_{|\mathcal{C}|}\} \).
Each satellite is equipped with wireless transceivers, either radio frequency or laser-based, enabling the dynamic establishment of satellite-to-ground links (SGLs) with ground users within these cells, as well as ISLs among the LEO satellites \cite{lai2023achieving, bakhsh2024multi}. 
Using spot beams, each cell can achieve bi-directional communication links. 
Given the non-uniform distribution of ground users, wireless communication services are concentrated in designated cells with active end-to-end communication demand, represented as \( \mathcal{C}_{active} \subseteq \mathcal{C} \). 
Unlike the conventional all-to-all service model, our analysis focuses on the impact of node failures within \( \mathcal{C}_{\text{active}} \). Thus, we can focus on an examination of how failures disrupt essential service areas.

\textbf{End-to-end Service Model.} In this study, we consider that end-to-end service demand arises only between ground cell pairs within \( \mathcal{C}_{\text{active}} \). 
We represent all possible end-to-end cell pairs using a two-dimensional array $\mathcal{I} \in \{0, 1\}^{|\mathcal{C}_{\text{active}}| \times |\mathcal{C}_{\text{active}}|}$, where each element corresponds to a binary indicator  \( \mathcal{I}_{ij} \) for cells $\mathbb{C}_i$ and $ \mathbb{C}_j$ in \( \mathcal{C}_{\text{active}}\). Specifically,  \( \mathcal{I}_{ij} = 1 \) signifies an active service demand between cells $\mathbb{C}_i$ and $ \mathbb{C}_j$, whereas \( \mathcal{I}_{ij} = 0 \) indicates no demand. For instance, \( \mathcal{I}_{ij} = 1, \forall  \mathcal{I}_{ij} \in \mathcal{I}\) indicates that there is end-to-end communication demand between every pair of active ground cells. This model allows us to represent and analyze the specific end-to-end wireless communication services impacted by node failures, offering insights into the resilience of designated services.

\textbf{Networking Model.}
Due to the rapid movement of LEO satellites relative to ground cells, with orbital periods typically ranging from approximately 90 minutes at 500 km altitude to 115 minutes at 1500 km altitude, both SGLs and ISLs change dynamically.
Wireless communication between nodes is only possible when they are within communication range. Once nodes move out of coverage, links are disrupted. 
These dynamic communication environments necessitate adaptive routing mechanisms.
To address these dynamics, the networking model incorporates contact graph routing (CGR) to facilitate dynamic path establishment between cells \cite{araniti2015contact}. CGR identifies the shortest delay paths within discrete time intervals, dynamically selecting routes based on the predicted availability of communication links. This approach ensures seamless and adaptive path establishment, effectively accommodating the periodic connectivity changes inherent in LEO satellite networks. Specifically, CGR operates based on a contact plan, a time-ordered list of predicted communication opportunities. Each contact is represented as \([t_s, t_e, u, v]\), indicating that there is a communication opportunity between \(u\) and \(v\) during the time interval  \([t_s, t_e]\). Here, $u$ and $v$ can be either a satellite or a ground cell. By using the contact plan,  temporal graphs can be constructed, upon which CGR calculates the shortest delay paths for designated services.

\vspace{-3 mm}
\subsection{Temporal Graph Model for Mega-Satellite Network}

Under conditions without satellite node failures, the evolution of the satellite network over the time period \(\mathcal{T} = [0, T]\) can be modeled as a continuous temporal graph \(G(t) = \{\mathcal{V}, E(t)\}, 0 \leq t \leq T\). Here, \(\mathcal{V} = \mathcal{S} \cup \mathcal{C}\) denotes the set of vertices, comprising both LEO satellites and ground cells, and \(E(t)\) represents the dynamically changing SGLs and ISLs. Each edge in \(E(t)\) is time-dependent and corresponds to a contact denoted by \([t_s, t_e, u, v]\), which represents a communication opportunity between vertices \(u\) and \(v\) from time \(t_s\) to \(t_e\). 
While the continuous temporal graph effectively captures the topological dynamics of the satellite network, the varying durations of communication links obstruct the practical application of routing algorithms for path selection. For instance, let us consider a shortest routing path calculated by CGR, comprising \(\mathbb{C}_1 \rightarrow \mathbb{S}_1 \rightarrow \mathbb{S}_2 \rightarrow \mathbb{C}_2\), with three edges/contacts
\([t_{s_1}, t_{e_1}, \mathbb{C}_1,  \mathbb{S}_1]\), \([t_{s_2}, t_{e_2}, \mathbb{S}_1,  \mathbb{S}_2]\) and \([t_{s_3}, t_{e_3}, \mathbb{S}_2,  \mathbb{C}_2]\). 
Given the durations \([t_{s_1}, t_{e_1}] = [0, 15]\) s, \([t_{s_2}, t_{e_2}] = [5, 15]\) s, and \([t_{s_3}, t_{e_3}] = [10, 20]\) s, the routing path is stable only during the overlapping time interval \([10, 15]\) s. However, the continuous temporal graph fails to capture the misalignment of contact times, potentially resulting in the assignment of time-misaligned or disconnected paths to ground cells, which impedes data transmission. 
Moreover, the extensive number of satellites in the LEO mega-constellation results in highly frequent communication link changes, making the computation of stable end-to-end service paths for real-time data communication highly challenging \cite{guo2024time}. 
To support stable service path establishment, we use a discrete temporal graph model, following the approach used in the Iridium system \cite{evans1998satellite}. 
This method discretizes time into small time intervals, effectively managing misaligned communication windows and facilitating the identification of stable service paths within specific time windows, thereby effectively enhancing CGR.

\textbf{Discrete Temporal Graph Model}.  
Based on the continuous temporal graph, we utilize a virtual topology-based approach that discretizes time into small time windows. Within each time window, the satellite network is considered static and stable. Specifically, the time discretization method divides the entire time horizon based on the \textit{contact plan} \cite{araniti2015contact}, which records all topology changes and resource variations caused by the periodic motion of satellites. Each entry in the contact plan, called a \textit{contact}, is denoted as a tuple \(<t_s, t_e, \mathbb{V}_i, \mathbb{V}_j >\), representing a communication opportunity from \(\mathbb{V}_i\) to \(\mathbb{V}_j\) during the time interval \([t_s, t_e]\), where  \(\mathbb{V}_i, \mathbb{V}_j \in \mathcal{V} = \mathcal{S} \cup \mathcal{C}\). 
It is worth noting that our method does not assume any grid-like satellite arrangement. Instead, it directly utilizes the time-varying positions and link distances derived from the contact plan, which is applicable to real-world satellite constellations with non-uniform topologies.
The discretization process first collects all time moments recorded in the contact plan and arranges them in ascending order. Each pair of adjacent time moments then defines a fine-grained time window. However, in mega-satellite constellations characterized by a large number of satellites and rapidly evolving topologies, such fine-grained windows can be extremely short, often lasting only a few milliseconds. 
If every minor topology change is used to define a separate time window, the resulting frequency of path switching may exceed the capability of routing algorithms to respond in time. Therefore, to effectively manage this high temporal variability, a service duration threshold is introduced to merge excessively short time windows and control the granularity of time discretization, with the aim of ensuring path stability. This threshold should be no shorter than the required service duration specified by user applications and no longer than the typical satellite-to-ground contact time, which is generally under 15 minutes in LEO satellite networks. In our simulations, we set this threshold to 60 seconds, as each LEO satellite typically covers a ground cell for tens of seconds up to several minutes in our cell-based model. This setting balances decision responsiveness and path stability and can be applied to general LEO constellations, with flexibility to adjust according to application needs and link dynamics.
Each resulting time window corresponds to a distinct virtual topology, forming a sequence of snapshot graphs \(\{\mathcal{G}_{\tau_1}, \mathcal{G}_{\tau_2}, \mathcal{G}_{\tau_3}, \ldots, \mathcal{G}_{\tau_n}\}\). 
\textit{Such a modeling process captures the heterogeneous spatial and temporal characteristics of satellite networks, including the dense connectivity in polar regions and sparse opportunities near inter-orbit seams.}
For clarity, the temporal graph discretization algorithm is detailed in \textbf{Algorithm \ref{temporal-graph-discretization}}. It initializes empty edge sets for each time window and iteratively assigns edges from the continuous graph to the corresponding snapshot graphs if their active periods overlap with a time window. This approach improves path stability and reduces the frequency of path switching.

\begin{algorithm}[t]
\caption{Temporal graph discretization algorithm}
\label{temporal-graph-discretization}
\begin{algorithmic}[1]
\item \textbf{Input:} The continuous temporal graph $G(t) = \{V, E(t)\}$, and the divided time windows $\{\tau_1, \tau_2, \tau_3, \ldots\}$. 
\item \textbf{Output:} The sequential series of discrete temporal graphs $\mathcal{G} =  \{\mathcal{G}_{\tau_1}, \mathcal{G}_{\tau_2}, \mathcal{G}_{\tau_3}, \ldots\}$, where $\mathcal{G}_{\tau_x} = \{V, E_{\tau_x}\}$.
\item \textbf{Initialize:} The empty set $\mathcal{G} = \{ \}$. 
\item \textbf{for} each time window $\tau_x = [t_{i-1}, t_i]$ \textbf{do}
\item \ \ \ \ Initialize $E_{\tau_x}$ as an empty edge set.
\item \ \ \ \ \textbf{for} each contact $[t_s, t_e, u, v]$ in $E(t)$ \textbf{do}
\item \ \ \ \ \ \ \ \ \textbf{if} $t_s \leq t_{i-1}$ and $t_e \geq t_i$ \textbf{then}
\item \ \ \ \ \ \ \ \ \ \ \ \ Add edge $(u, v)$ to $E_{\tau_x}$.
\item \ \ \ \ Create discrete temporal graph $\mathcal{G}_{\tau_x} = \{V, E_{\tau_x}\}$.
\item \ \ \ \ Update $\mathcal{G}  \longleftarrow \mathcal{G} \cup \{ \mathcal{G}_{\tau_x} \}$.
\item \textbf{return} $\mathcal{G}  = \{\mathcal{G}_{\tau_1}, \mathcal{G}_{\tau_2}, \mathcal{G}_{\tau_3}, \ldots\}$.
\end{algorithmic}
\end{algorithm}

At each time window \(\tau_x\), the network snapshot is represented as a directed graph \(\mathcal{G}_{\tau_x} = (V, E_{\tau_x})\), where $V$ denotes the set of nodes, and $E_{\tau_x}$  is the set of possible communication links during time window $\tau_x$. 
Each link \(e_{\tau_x} = (u, v) \in E_{\tau_x}\) represents a possible communication link between vertex \(u\) and \(v\) within the time window \(\tau_x\), with a propagation delay \( L^{\tau_x}_{u, v} \).
Additionally, the availability of each vertex \(v \in V\) during time window \(\tau_x\) is indicated by a binary indicator \(A_{v, \tau_x} \in \{0, 1\}\), where \(A_{v, \tau_x} = 1\) if vertex \(v\) is available, and \(A_{v, \tau_x} = 0\) otherwise. Similarly, the availability of each link \(e_{\tau_x} \in E_{\tau_x}\) is indicated by a binary indicator \(J_{e_{\tau_x}}\), where \(J_{e_{\tau_x}} = 1\) indicates that the link  \(e_{\tau_x} \) is available during time window \(\tau_x\), and \(J_{e_{\tau_x}} = 0\) otherwise.

To maintain consistency, the availability of a vertex and its associated links must adhere to the following dependency. If vertex \(v\) is unavailable, i.e., \(A_{v, \tau_x} = 0\), then all links connected to \(v\) in \(\mathcal{G}_{\tau_x}\) including both incoming and outgoing links, are also unavailable. This relationship is formalized as:
\begin{equation}
     \text{if } A_{v, \tau_x} = 0,  \text{ then } J_{e_{\tau_x}} = 0, \forall e_{\tau_x} = (u, v) \text{ or } (v, w) \in E_{\tau_x}. 
\end{equation}
This formalization of node and link availability dependencies provides the foundation for analyzing network resilience under dynamic conditions. It allows us to model and evaluate the networking performance of the satellite network in response to changes in node and link availability.

\textbf{Path Model.} Leveraging the discrete temporal graph model, we are able to compute paths between ground cells across different time windows. Specifically, we define the shortest path between cell \(\mathbb{C}_1\) and cell \(\mathbb{C}_2\) during time window \(\tau\) as a set \(p^{\tau}_{\mathbb{C}_1 \rightarrow \mathbb{C}_2}\). The shortest paths between all pairs of cells can be efficiently computed using standard shortest path algorithms. Under normal conditions, the vertex set \(V = \mathcal{S} \cup \mathcal{U}\) and the shortest path in each  discrete temporal graph remains constant.  However, when a node failure occurs, the affected nodes are removed from the temporal graph, causing corresponding disruptions in the links and paths. Therefore, we will define the node failure event model in the next subsection.

\vspace{-3 mm}

\subsection{Node Failure Event Model}

The space environment presents various risk factors that can lead to node failures, such as solar storms, cosmic debris collisions, extreme temperature variations and hardware malfunctions. These events affect both the vertices and edges of the discrete temporal graph.
To accurately model these node failure events, we define a node failure event and analyze its impact based on the discrete temporal graph.

\begin{definition}[Node Failure Event] A node failure event is defined as a tuple \( F = (t_f, \mathcal{S}_f) \), where \( t_f \) denotes the timestamp of the failure event, and \( \mathcal{S}_f \subseteq \mathcal{S} \) represents the set of satellite nodes that fail at time \( t_f \).  
For each satellite \( \mathbb{S}_{\sigma} \in \mathcal{S}_f \), we denote its failure by \( F(t_f, \mathbb{S}_{\sigma}) = 1 \). 
The occurrence of this failure event modifies the network topology starting from the time window \(\tau_f\) that includes \(t_f\), impacting  all subsequent temporal graphs \(\{\mathcal{G}_{\tau_f}, \mathcal{G}_{\tau_{f+1}}, \mathcal{G}_{\tau_{f+2}}, \ldots, \mathcal{G}_{\tau_n}\}\). The cardinality \(|\mathcal{S}_f|\) represents the number of failed satellite nodes, with \(|\mathcal{S}_f| = 1\) indicating a single node failure and \(|\mathcal{S}_f| > 1\) indicating multiple node failures. 
\end{definition}

A node failure event results in two primary outcomes:

\begin{itemize}
    \item \textbf{Edge Removal:} All edges associated with the failed nodes are removed from the affected temporal graphs.
   
    \item \textbf{Path Recalculation:} 
    Any shortest path in the affected temporal graphs that passes through a failed node must be recalculated.

\end{itemize}
\textit{It is worth noting that, since the time windows are determined during the initial step of temporal graph construction, node failures occurring after this step will not trigger a re-partitioning of the time windows. Instead, they only affect the snapshot graphs within the predefined time windows by altering network connectivity and available link resources.}

Thus, for each time window \(\tau_x = [t_x, t_{x+1}]\), if \(t_{x+1} \geq t_f\), the node failure event modifies the network topology, resulting in an updated discrete temporal graph \(\mathcal{G}'_{\tau_x} = (V'_{\tau_x}, E'_{\tau_x})\), where \(V'_{\tau_x}\) and \(E'_{\tau_x}\) denote the residual vertex and edge sets, respectively. These sets are defined as follows:
\begin{equation}
V'_{\tau_x} = V \setminus \mathcal{S}_f,
\end{equation}
\begin{equation}
E^{'}_{\tau_x} = \{(u, v) \in E_{\tau_x} \mid u \notin \mathcal{S}_f \text{ and } v \notin \mathcal{S}_f\}.
\end{equation}
Here, \(V'_{\tau_x}\) and \(E'_{\tau_x}\) represent the remaining vertices and edges after removing those associated with failed nodes.

To maintain networking functionality and service connectivity, the shortest paths between active cell pairs \(\mathbb{C}_i \) and \(\mathbb{C}_j \) must be recalculated if they are impacted by node failures. Specifically, $\forall \mathbb{C}_i, \mathbb{C}_j \in \mathcal{C}_{\text{active}}$, if \(\mathcal{I}_{ij} = 1\) and the original shortest path \(p_{\mathbb{C}_i \rightarrow \mathbb{C}_j}^{\tau_x} \) passes through any node in \(\mathcal{S}_f \), then a new shortest path, denoted as \({p'}_{\mathbb{C}_i \rightarrow \mathbb{C}_j}^{\tau_x}\), must be computed in the updated graph \(\mathcal{G}^{'}_{\tau_x} \). Formally, this can be expressed as follows: $\forall \mathbb{C}_i, \mathbb{C}_j \in \mathcal{C}_{\text{active}}, \tau_x = [t_x, t_{x+1}], t_{x+1} \geq t_f$, 
\begin{equation}
{p}_{\mathbb{C}_i \rightarrow \mathbb{C}_j}^{\tau_x} \longleftarrow 
\begin{cases}
{p'}_{\mathbb{C}_i \rightarrow \mathbb{C}_j}^{\tau_x}, & \text{if } \mathcal{I}_{ij} = 1 \text{ and } {p}_{\mathbb{C}_i \rightarrow \mathbb{C}_j}^{\tau_x} \cap \mathcal{S}_f \neq \emptyset, \\
{p}_{\mathbb{C}_i \rightarrow \mathbb{C}_j}^{\tau_x}, & \text{otherwise}.
\end{cases}
\end{equation}
Here, \( {p}_{\mathbb{C}_i \rightarrow \mathbb{C}_j}^{\tau_x} \cap \mathcal{S}_f \neq \emptyset \) indicates that the original shortest path \( {p}_{\mathbb{C}_i \rightarrow \mathbb{C}_j}^{\tau_x} \) traverses at least one failed node within \( \mathcal{S}_f \).
By modeling node failures and their impacts on the discrete temporal graph, we can effectively account for the dynamic evolution and resilience of mega-satellite constellations.

\vspace{-3 mm}

\subsection{Problem Statement}

\begin{figure}[t!]
	\centering
 	\includegraphics[width=66mm]{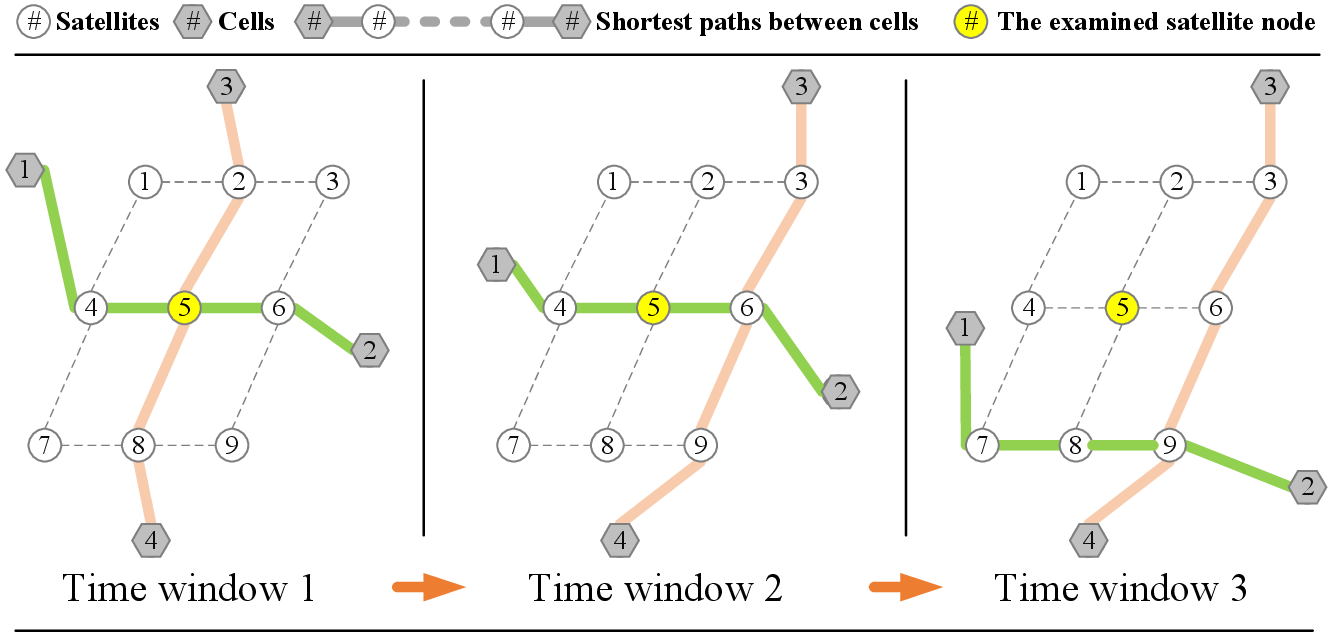}\\
	\caption{The evolution of critical nodes in LEO mage-satellite constellations with inter-satellite links (ISLs) for end-to-end networking. }
\label{fig:key_node_evolution}
% \vspace{-3 mm}
\end{figure}

In satellite network systems, data communication between ground cells relies on shortest path routing, which dynamically adjusts paths across time windows to account for periodic link disruptions, as dictated by the CGR strategy. Additionally, when node failures occur, shortest paths must further adapt, introducing a complex interplay between time-varying link interruptions and node failures that complicates failure analysis. 
As illustrated in Figure \ref{fig:key_node_evolution}, at time window \(\tau_1\), satellite node \(\mathbb{S}_5\) is a crucial relay node in the shortest paths between cell \(\mathbb{C}_1\) and cell \(\mathbb{C}_2\), and between cell \(\mathbb{C}_3\) and cell \(\mathbb{C}_4\). At time window \(\tau_2\), \(\mathbb{S}_5\) continues to serve as a relay for the path between cell \(\mathbb{C}_1\) and cell \(\mathbb{C}_2\) but is no longer involved in the path between cell \(\mathbb{C}_3\) and cell \(\mathbb{C}_4\). At time window \(\tau_3\), \(\mathbb{S}_5\) no longer serves as a relay in any shortest paths, rendering it idle.
The impact of node failures is therefore time-dependent. For instance, if node \(\mathbb{S}_5\) fails at \(\tau_1\), communication along the shortest paths between both cell pairs \(\mathbb{C}_1\)-\(\mathbb{C}_2\) and \(\mathbb{C}_3\)-\(\mathbb{C}_4\) would be disrupted, resulting in degraded connectivity and possibly increased path delays. However, a failure of \(\mathbb{S}_5\) at \(\tau_2\) would only affect the path between cell \(\mathbb{C}_1\) and cell \(\mathbb{C}_2\). At time window \(\tau_3\), a failure of \(\mathbb{S}_5\) would have no effect on any shortest paths, as it would no longer be part of them.
While existing methods primarily focus on developing resilient routing and path recovery strategies, they have not delved into the evolving roles and importance of individual nodes over time. 
This gap limits the understanding of how node failures dynamically affect satellite networks, potentially hindering the deployment of more effective resilience strategies.
Therefore, there is a critical need within the network research community for systematic and effective methods to analyze the evolution of critical nodes, and assess the time-dependent and service-related impact of node failures.  This analysis is crucial for guiding the design, deployment and maintenance of next-generation satellite networks, as it enables robust performance evaluation, failure mitigation, and precise impact assessment.

In this study, we aim to systematically examine the temporal evolution of critical satellites and assess the impact of both single and multiple node failures on inter-satellite networking for designated end-to-end wireless communication services. Our investigation centers on three critical research questions:

\begin{enumerate} 

\item \textbf{Temporal evolution of node importance:} 
Given specific active cells and end-to-end service demands, how does the importance of individual satellite nodes vary over time in response to regular satellite network dynamics?

\item \textbf{Impact of node failures on inter-satellite networking over time:} What are the immediate and subsequent effects on inter-satellite networking when and after the node failure event occurs, particularly in terms of service connectivity and path delay?

 \item \textbf{Resilience capability of mega-satellite constellations:} To what extent can the mega-satellite constellations tolerate node failures before experiencing a complete disruption of service, and what is the associated recovery strategy for network restoration? 

\end{enumerate}

By addressing these questions, we provide insights into node dynamics and resilience in mega-satellite constellations, helping optimize routing and enhance system robustness. To achieve this, we will develop the critical node analysis framework, define the quantitative metric, and design corresponding algorithms for node failure impact analysis.

\vspace{ -3mm}

\section{Service-Aware Temporal Betweenness-Based Node Failure Impact Analysis Framework}  \label{sec:framework}

In this section, we propose a service-aware temporal betweenness (SATB)-based node failure impact analysis framework. First, we briefly review traditional betweenness centrality and its limitations. Then, we introduce the SATB metric extended from the traditional betweenness centrality for mega-satellite constellations. Next, we detail how to identify and rank critical nodes using the SATB. Finally, we propose algorithms to evaluate node failure impacts on inter-satellite networking across current and future time windows.

\vspace{-3 mm}
\subsection{Preliminaries on The Traditional Betweenness Centrality}

Betweenness centrality is a widely used metric for quantifying the importance of a node in a network by measuring how often a node lies on the shortest paths between all pairs of nodes in a network. For a node \( v \) in a static graph $G = (V, E)$, its betweenness centrality is formally defined as:
\begin{equation}
\label{equation5}
    B(v) = \sum_{\substack{s, d, v \in V \\ s \neq d \neq v}} \frac{\sigma_{s,d}(v)}{\sigma_{s,d}},
\end{equation}
where:
\begin{itemize}
    \item  \( \sigma_{s,d} \) is the total number of shortest paths between source node \( s \) and destination node \( d \), \( \sigma_{s,d} \geq 0, \sigma_{s,d} \in \mathbb{N} \),
   \item \( \sigma_{s,d}(v) \) is the number of shortest paths from \( s \) to \( d \) that pass through node \( v \).
\end{itemize}

Equation \ref{equation5} sums up the ratio of the number of shortest paths passing through node \( v \) to the total number of shortest paths between all pairs of nodes \( s \) and \( d \), where \( s \neq v \neq  d\). The betweenness centrality thus reflects the fraction of shortest paths in the network that pass through a given node, highlighting its role in facilitating communication among nodes.

In static networks, the betweenness centrality is a powerful tool for identifying critical nodes that control information flow. However, in dynamic or time-varying networks, this metric must be adapted to account for temporal changes in the network topology. This involves recalculating shortest paths as the network evolves, making it more suitable for assessing the dynamic role of nodes in real-time communication systems such as satellite networks. 
Especially, when applied to dynamic and service-oriented satellite networks, several key limitations of classic betweenness centrality arise:

\begin{itemize}
    \item  \textbf{All-to-all service assumption}: 
    The traditional betweenness centrality assumes an all-to-all service model, which is impractical in satellite systems where only active node pairs require connections at a given time. This neglects the non-uniform and on-demand nature of connectivity in real-world scenarios.

    \item \textbf{Assumption on non-unique shortest paths}: 
    The traditional betweenness centrality assesses node importance based on all shortest paths, emphasizing topological structure. In satellite networks, however, only one shortest path is typically used for an established end-to-end service, making it misaligned with real-world path usage.

    \item \textbf{Fixed topology assumption}:  The traditional betweenness centrality is inherently designed for static networks and does not account for dynamic changes in node availability or network topology. In dynamic LEO satellite networks, the importance of nodes may vary over time due to changes in the constellation or communication conditions. This renders the traditional betweenness centrality insufficient for capturing the temporal variations in node centrality in dynamic  satellite networks.
    
\end{itemize}

To represent node importance in dynamic, service-oriented satellite networks, we adapt the traditional betweenness centrality to a time-varying and service-specific context, enabling assessment of node failure impacts over time.

\vspace{-3mm}
\subsection{The Proposed Service-Aware Temporal Betweenness Metric for Mega-Satellite Constellations}

While betweenness centrality captures the number of shortest paths passing through a node, representing its topological importance, it is not fully applicable to dynamically evolving satellite networks with uneven service distribution. Specifically, traditional betweenness centrality fails to capture time-varying importance and service-related importance, as it relies on all-pairs shortest paths, assuming service demand exists between every cell pair. However, this assumption does not always hold, particularly in satellite networks, where communication occurs only between certain active cell pairs based on service demands \cite{lai2023achieving}. To address these limitations, we introduce the service-aware temporal betweenness (SATB) metric, which integrates two essential aspects corresponding to the aforementioned features of node failure impact:
\begin{itemize}
    \item \textbf{Time-dependence}: Satellite networks are inherently dynamic, with node connectivity changing over time. We redefine the original betweenness centrality, which is initially represented as a single value, as a time series, with each value representing a node's importance within a specific time window, thereby capturing temporal variations in the satellite network topology.

    \item \textbf{Service-awareness}: In real-world scenarios, only active ground cells have end-to-end service demand at a given time, resulting in communication between specific source-destination pairs. By focusing on these active pairs rather than all-to-all cell pairs, our metric more accurately reflects real-world satellite networks.
\end{itemize}

Therefore, we formally define the SATB metric as follows:
\begin{definition}[Service-aware temporal betweenness (SATB)]
The service-aware temporal betweenness \(\mathcal{B}_{\mathbb{S}_v}\) for a satellite node \(\mathbb{S}_v \in \mathcal{S}\) is a time series and a vector representing the importance across discrete time windows, which is defined as:
\begin{equation}
    \mathcal{B}_{\mathbb{S}_v} = [\beta^{\tau_1}_{\mathbb{S}_v}, \beta^{\tau_2}_{\mathbb{S}_v}, \beta^{\tau_3}_{\mathbb{S}_v}, \ldots, \beta^{\tau_n}_{\mathbb{S}_v} ],
\end{equation}
where each component \(\beta^{\tau_x}_{\mathbb{S}_v}\) represents the SATB value of node \(\mathbb{S}_v\) in time window \(\tau_x\). The SATB component \(\beta^{\tau_x}_{\mathbb{S}_v}\) for a time window \(\tau_x\) is defined as the total count of shortest delay paths passing through \(\mathbb{S}_v\) between all active source-destination pairs \((\mathbb{C}_i, \mathbb{C}_j)\) with end-to-end service demand, where only one path is considered per source-destination pair. 
Formally, we define
\begin{equation}
    \beta^{\tau_x}_{\mathbb{S}_v} = \sum_{\substack{(\mathbb{C}_i, \mathbb{C}_j) \in \mathcal{C}_{\text{active}} \times \mathcal{C}_{\text{active}} \\ \mathcal{I}_{ij} = 1}} \delta^{\tau_x}_{\mathbb{C}_i \rightarrow \mathbb{C}_j}(\mathbb{S}_v)
\end{equation}
where:
\begin{itemize}
    \item  \(\delta^{\tau_x}_{\mathbb{C}_i \rightarrow \mathbb{C}_j}(\mathbb{S}_v) \in \{0, 1\} \), and \(\delta^{\tau_x}_{\mathbb{C}_i \rightarrow \mathbb{C}_j}(\mathbb{S}_v) = 1\) if the shortest path between \(\mathbb{C}_i\) and \(\mathbb{C}_j\) passes through satellite node \(\mathbb{S}_v\) during time window \(\tau_x\), and \(0\) otherwise;
    \item \(\mathcal{I}_{ij} = 1\) indicates that there is service demand between active ground cells \(\mathbb{C}_i\) and \(\mathbb{C}_j  \in  \mathcal{C}_{\text{active}} \).
\end{itemize}

\end{definition}

Based on the above definition, the following fundamental lemmas can be derived:

\begin{lemma}[Range of SATB Values]
For any satellite node \(\mathbb{S}_v\) within a given time window \(\tau\), the SATB component \(\beta^{\tau}_{\mathbb{S}_v}\) is a non-negative integer satisfying \(0 \leq \beta^{\tau}_{\mathbb{S}_v} \leq \sum_{\mathbb{C}_i, \mathbb{C}_j \in \mathcal{C}_{\text{active}}} \mathcal{I}_{ij}\), where \(\mathcal{I}_{ij} =1 \) indicates the presence of a service demand between active ground cells \(\mathbb{C}_i\) and \(\mathbb{C}_j \in  \mathcal{C}_{\text{active}}\).
\end{lemma}

\begin{proof}
The SATB component \(\beta^{\tau}_{\mathbb{S}_v}\) measures the count of shortest paths that pass through \(\mathbb{S}_v\) for each active source-destination pair within the time window \(\tau\). By definition, \(\beta^{\tau}_{\mathbb{S}_v}\) is a non-negative integer, as it counts these paths. The maximum possible value for \(\beta^{\tau}_{\mathbb{S}_v}\) occurs when all shortest paths between active cell pairs pass through \(\mathbb{S}_v\), yielding \(\beta^{\tau}_{\mathbb{S}_v} \leq \sum_{\mathbb{C}_i, \mathbb{C}_j \in \mathcal{C}_{\text{active}}} \mathcal{I}_{ij}\).
\end{proof}

\begin{lemma}[SATB Value for Non-Participating Nodes]
If a satellite node \(\mathbb{S}_v\) does not participate in routing for any active source-destination pairs within a time window \(\tau\), then \(\beta^{\tau}_{\mathbb{S}_v} = 0\).
\end{lemma}

\begin{proof}
When node \(\mathbb{S}_v\) does not belong to the shortest paths between any active source-destination pairs during the time window \(\tau\), no shortest paths contribute to \(\beta^{\tau}_{\mathbb{S}_v}\). Consequently, \(\beta^{\tau}_{\mathbb{S}_v} = 0\) by the definition of SATB.
\end{proof}

\begin{lemma}[Node Failure Impact on SATB]
If a satellite node \(\mathbb{S}_v\) fails at time window \(\tau\), then the SATB component \(\beta^{\tau'}_{\mathbb{S}_v} = 0\) for all time windows \(\tau' \geq \tau\).
\end{lemma}

\begin{proof}
Upon satellite node failure, \(\mathbb{S}_v\) ceases to be available for routing from the current time window onward. As a result, \(\mathbb{S}_v\) cannot appear in any shortest paths for future time windows, leading to \(\beta^{\tau'}_{\mathbb{S}_v} = 0\) for all \(\tau' \geq \tau\).
\end{proof}

Furthermore, the configuration of time window length can also affect both satellite importance and the impact of node failures. Smaller time windows tend to yield higher importance values and shorter failure durations, while larger time windows may lead to lower importance and longer failure impact, as routing decisions are made on a per time window basis. However, if the time window is too small, path switching may become too frequent or even infeasible. Therefore, the time window size must be set based on the service duration and network responsiveness, as discussed in \cite{guo2024enhanced}.

\subsubsection{Algorithm for calculating SATB for a single satellite}
For clarity, the detailed steps for computing the SATB vector for a given satellite node \({\mathbb{S}_v}\) are outlined in \textbf{Algorithm \ref{temporal-betweenness}}. This procedure captures both the dynamic topology and designated wireless communication services in satellite networks, by processing temporal graphs and active source-destination ground cell pairs to quantify the node's SATB over time. The algorithm takes as input the discrete temporal graphs \(\{\mathcal{G}_{\tau_1}, \mathcal{G}_{\tau_2}, \ldots, \mathcal{G}_{\tau_n}\}\), the set of active source-destination cell pairs \(\mathcal{H} = \{ (\mathbb{C}_i, \mathbb{C}_j) \mid \mathbb{C}_i, \mathbb{C}_j \in \mathcal{C}_{\text{active}}, \mathcal{I}_{ij} = 1 \}\), and the target satellite \(\mathbb{S}_v\). The active cell pairs can be flexibly configured by setting the binary indicator \(\mathcal{I}_{ij} \in \{0, 1\}\) based on practical service demands. For example, if satellite-based communication between cell pairs such as Xi' an and Singapore is required, we set \(\mathcal{I}_{ij} = 1\); otherwise, \(\mathcal{I}_{ij} = 0\). In real-world deployments, these values can be flexibly configured by satellite operators such as Starlink in accordance with service-level requirements. In our simulation, we set \(\mathcal{I}_{ij} = 1\) for all \(\mathbb{C}_i, \mathbb{C}_j \in \mathcal{C}_{\text{active}}\) to evaluate the temporal variation in SATB for critical satellites under full-demand scenarios. The algorithm begins by initializing an empty vector \(\mathcal{B}_{\mathbb{S}_v}\) to store SATB values for \(\mathbb{S}_v\) across all time windows. For each time window \(\tau_x\), the SATB value \(\beta^{\tau_x}_{\mathbb{S}_v}\) is initialized as zero. Then, for each active cell pair \((\mathbb{C}_i, \mathbb{C}_j)\), the algorithm computes the shortest path \(p^{\tau_x}_{\mathbb{C}_i \rightarrow \mathbb{C}_j}\) in \(\mathcal{G}_{\tau_x}\). If the target satellite \(\mathbb{S}_v\) lies on this path, \(\beta^{\tau_x}_{\mathbb{S}_v}\) is incremented by 1. After processing all pairs, \(\beta^{\tau_x}_{\mathbb{S}_v}\) is appended to \(\mathcal{B}_{\mathbb{S}_v}\). The algorithm ultimately returns the SATB vector \(\mathcal{B}_{\mathbb{S}_v}\), which yields a temporal profile of the satellite’s evolving importance throughout the observation period.

\begin{algorithm}[t]
\caption{Computation of service-aware temporal betweenness (SATB) for a single satellite}
\label{temporal-betweenness}
\begin{algorithmic}[1]
\item \textbf{Input:} The discrete temporal graphs \(\{\mathcal{G}_{\tau_1}, \mathcal{G}_{\tau_2}, \ldots, \mathcal{G}_{\tau_n}\}\), the set of active source-destination cell pairs \( \mathcal{H} = \{    (\mathbb{C}_i, \mathbb{C}_j) | \mathbb{C}_i, \mathbb{C}_j \in \mathcal{C}_{\text{active}}, \mathcal{I}_{ij} = 1 \}\), and satellite \({\mathbb{S}_v}\).
\item \textbf{Output:} The SATB vector \(\mathcal{B}_{\mathbb{S}_v} = [\beta^{\tau_1}_{\mathbb{S}_v}, \beta^{\tau_2}_{\mathbb{S}_v}, \beta^{\tau_3}_{\mathbb{S}_v}, \ldots, \beta^{\tau_n}_{\mathbb{S}_v}]\).
\item Initialize an empty vector \(\mathcal{B}_{\mathbb{S}_v}\) to store the SATB  values.
\item \textbf{for} each time window \(\tau_x \in \{{\tau_1}, {\tau_2}, \ldots,{\tau_n}\} \) \textbf{do}
    \item \ \ \ Initialize \(\beta^{\tau_x}_{\mathbb{S}_v} = 0\).
    \item \ \ \ \textbf{for} each source-destination cell pair \((\mathbb{C}_{i},\mathbb{C}_{j}) \in \mathcal{H} \)  \textbf{do}
       \item \ \ \ \ \ \ Find the shortest path \(p^{\tau_x}_{\mathbb{C}_{i} \rightarrow \mathbb{C}_{j}}\) in \(\mathcal{G}_{\tau_x}\).
        \item \ \ \ \ \ \ \textbf{if} \(\mathbb{S}_v\) lies on \(p^{\tau_x}_{\mathbb{C}_{i} \rightarrow \mathbb{C}_{j}}\) \textbf{then}
             \item \ \ \ \ \ \ \ \ \ Update \(\beta^{\tau_x}_{\mathbb{S}_v} = \beta^{\tau_x}_{\mathbb{S}_v} +1 \).
    \item \ \ \ Append \(\beta^{\tau_x}_{\mathbb{S}_v}\) to \(\mathcal{B}_{\mathbb{S}_v}\).
\item \textbf{return}  \(\mathcal{B}_{\mathbb{S}_v} = [\beta^{\tau_1}_{\mathbb{S}_v}, \beta^{\tau_2}_{\mathbb{S}_v}, \beta^{\tau_3}_{\mathbb{S}_v}, \ldots, \beta^{\tau_n}_{\mathbb{S}_v}]\). 
\end{algorithmic}
\end{algorithm}

% \vspace{ -3mm}

\subsubsection{Algorithm for calculating SATB once for all satellites}
To enhance computational efficiency, SATB values for all satellite nodes can be computed simultaneously, reducing complexity to the same level as calculating SATB for a single satellite. This approach leverages collective evaluation within each time window, effectively eliminating redundant computations. \textbf{Algorithm \ref{computation-SATB-all}} provides the detailed steps to the efficient computation of SATB vectors. In each time window \( \tau_x \), the algorithm begins by initializing a temporary list \( \mathcal{L} \) to store intermediary satellite nodes encountered on shortest paths between active source-destination cell pairs. For each pair \( (\mathbb{C}_i, \mathbb{C}_j) \in \mathcal{H} \), the shortest path \( p^{\tau_x}_{\mathbb{C}_{i} \rightarrow \mathbb{C}_{j}} \) is identified within the temporal graph \( \mathcal{G}_{\tau_x} \), and all intermediary satellite nodes on this path, excluding source and destination cells, are added to \( \mathcal{L} \).
Upon completing path computations for a given time window, the algorithm generates a frequency map \( D \), logging the count of each satellite node in \( \mathcal{L} \). This count, representing the node's frequency as an intermediary on shortest paths within the time window, is then assigned to the respective SATB value \( \beta^{\tau_x}_{\mathbb{S}_v} \). By aggregating these results, the algorithm efficiently yields a complete SATB profile \( \mathcal{B} \) for all satellites across the temporal sequence of graphs.

% \vspace{ -3mm}

\subsubsection{An example for calculating SATB}
Consider a satellite network with four satellites \(\mathcal{S} = \{\mathbb{S}_1, \mathbb{S}_2, \mathbb{S}_3, \mathbb{S}_4\}\) and four ground cells \(\mathcal{C} = \{\mathbb{C}_1, \mathbb{C}_2, \mathbb{C}_3, \mathbb{C}_4\}\). The fixed active source-destination pairs are \((\mathbb{C}_1, \mathbb{C}_3)\), \((\mathbb{C}_2, \mathbb{C}_4)\), and \((\mathbb{C}_1, \mathbb{C}_4)\) across three time windows \(\tau_1\), \(\tau_2\), and \(\tau_3\). In time window \(\tau_1\), the shortest paths are \(\mathbb{C}_1 \rightarrow \mathbb{S}_1 \rightarrow \mathbb{C}_3\) for \((\mathbb{C}_1, \mathbb{C}_3)\), \(\mathbb{C}_2 \rightarrow \mathbb{S}_2 \rightarrow \mathbb{C}_4\) for \((\mathbb{C}_2, \mathbb{C}_4)\), and \(\mathbb{C}_1 \rightarrow \mathbb{S}_1 \rightarrow \mathbb{S}_2 \rightarrow \mathbb{C}_4\) for \((\mathbb{C}_1, \mathbb{C}_4)\). Thus, the SATB values are \(\beta^{\tau_1}_{\mathbb{S}_1} = 2\), \(\beta^{\tau_1}_{\mathbb{S}_2} = 2\), \(\beta^{\tau_1}_{\mathbb{S}_3} = 0\), and \(\beta^{\tau_1}_{\mathbb{S}_4} = 0\). In time window \(\tau_2\), the shortest paths are \(\mathbb{C}_1 \rightarrow \mathbb{S}_3 \rightarrow \mathbb{C}_3\) for \((\mathbb{C}_1, \mathbb{C}_3)\), \(\mathbb{C}_2 \rightarrow \mathbb{S}_2 \rightarrow \mathbb{C}_4\) for \((\mathbb{C}_2, \mathbb{C}_4)\), and \(\mathbb{C}_1 \rightarrow \mathbb{S}_3 \rightarrow \mathbb{S}_4 \rightarrow \mathbb{C}_4\) for \((\mathbb{C}_1, \mathbb{C}_4)\). Consequently, \(\beta^{\tau_2}_{\mathbb{S}_1} = 0\), \(\beta^{\tau_2}_{\mathbb{S}_2} = 1\), \(\beta^{\tau_2}_{\mathbb{S}_3} = 2\), and \(\beta^{\tau_2}_{\mathbb{S}_4} = 1\). In time window \(\tau_3\), the shortest paths are \(\mathbb{C}_1 \rightarrow \mathbb{S}_4 \rightarrow \mathbb{C}_3\) for \((\mathbb{C}_1, \mathbb{C}_3)\), \(\mathbb{C}_2 \rightarrow \mathbb{S}_2 \rightarrow \mathbb{C}_4\) for \((\mathbb{C}_2, \mathbb{C}_4)\), and \(\mathbb{C}_1 \rightarrow \mathbb{S}_2 \rightarrow \mathbb{S}_4 \rightarrow \mathbb{C}_4\) for \((\mathbb{C}_1, \mathbb{C}_4)\). Thus, \(\beta^{\tau_3}_{\mathbb{S}_1} = 0\), \(\beta^{\tau_3}_{\mathbb{S}_2} = 2\), \(\beta^{\tau_3}_{\mathbb{S}_3} = 0\), and \(\beta^{\tau_3}_{\mathbb{S}_4} = 2\). Consequently, the SATB vectors for the four satellites are \(\mathcal{B}_{\mathbb{S}_1} = [2, 0, 0]\), \(\mathcal{B}_{\mathbb{S}_2} = [2, 1, 2]\), \(\mathcal{B}_{\mathbb{S}_3} = [0, 2, 0]\), and \(\mathcal{B}_{\mathbb{S}_4} = [0, 1, 2]\). This example demonstrates how the SATB values for the satellites vary across different time windows. 

\begin{algorithm}[t!]
\caption{Computation of SATB for all satellite nodes}
\label{computation-SATB-all}
\begin{algorithmic}[1]
\item \textbf{Input:} The discrete temporal graphs \(\{\mathcal{G}_{\tau_1}, \mathcal{G}_{\tau_2}, \ldots, \mathcal{G}_{\tau_n}\}\), the set of active source-destination cell pairs \( \mathcal{H} = \{    (\mathbb{C}_i, \mathbb{C}_j) | \mathbb{C}_i, \mathbb{C}_j \in \mathcal{C}_{\text{active}}, \mathcal{I}_{ij} = 1 \}\), and satellite set \( \mathcal{S} \).
\item \textbf{Output:} The SATB for all satellites \( \mathcal{B} = \{ \mathcal{B}_{\mathbb{S}_v} |   \mathbb{S}_v \in \mathcal{S}\} \).
\item Initialize the SATB vector \( \mathcal{B}_{\mathbb{S}_v} = \mathbf{0},  \forall \mathbb{S}_v \in \mathcal{S} \).
\item \textbf{for} each time window \( \tau_x \in \{\tau_1, \tau_2, \ldots, \tau_n\} \) \textbf{do}
 \item \ \ \  Initialize \( \mathcal{L} =\emptyset \) to store intermediate satellite nodes.
 \item \ \ \ \textbf{for} each source-destination cell pair \((\mathbb{C}_{i},\mathbb{C}_{j}) \in \mathcal{H} \)  \textbf{do}
       \item \ \ \ \ \ \ Find the shortest path \(p^{\tau_x}_{\mathbb{C}_{i} \rightarrow \mathbb{C}_{j}}\) in \(\mathcal{G}_{\tau_x}\).
         \item \ \ \ \ \ \ Update \(   \mathcal{L} = \mathcal{L} \cup \left\{ v \mid v \in p^{\tau_x}_{\mathbb{C}_i \to \mathbb{C}_j}, v \notin \{\mathbb{C}_i, \mathbb{C}_j\} \right\}  \).
\item \ \ \  Let \( D = \{ \mathbb{S}_v :  \text{freq}( \mathbb{S}_v) = \text{count of }  \mathbb{S}_v \text{ in } \mathcal{L} \mid \mathbb{S}_v  \in \mathcal{L} \} \). 
\item \ \ \ \textbf{for}  each satellite node \( \mathbb{S}_v \) in \( D \) \textbf{do}
   \item \ \ \ \ \ \  Update \(\beta^{\tau_x}_{\mathbb{S}_v} = \text{freq}( \mathbb{S}_v) \). 
\item \textbf{return} The SATB for all satellites \( \mathcal{B} = \{ \mathcal{B}_{\mathbb{S}_v} |   \mathbb{S}_v \in \mathcal{S}\} \).
\end{algorithmic}
\end{algorithm}

In summary, the SATB metric extends traditional betweenness centrality by incorporating temporal and service-specific factors, providing a more accurate assessment of node importance and offering insights for optimizing satellite network strategies and resilience.

\vspace{-5mm}

\subsection{Identification and Ranking of Critical Nodes}

In mega-satellite networks, \textit{critical nodes} are those that play vital roles in maintaining end-to-end communication between active cells. 
These nodes are identified based on their involvement in the shortest paths that connect active source-destination cell pairs during each time window. 
For a given time window \(\tau\), we define the set of critical nodes as those nodes \(\mathbb{S}_v \in 
 \mathcal{S}\) whose SATB values are greater than zero: 
\begin{equation}
    \mathcal{C}_{\tau} = \{ \mathbb{S}_v \in \mathcal{S} \mid \beta^{\tau}_{\mathbb{S}_v} > 0 \},
\end{equation}
where \(\mathcal{C}_{\tau}\) represents the set of critical satellites that actively facilitate communication during time window \(\tau\). Additionally, the ranking indicator \(I^{\tau}_{\mathbb{S}_v}\) is introduced to represent the relative importance of node \(\mathbb{S}_v\) within the set of critical nodes for time window \(\tau\), where \(I^{\tau}_{\mathbb{S}_v} > 0\) and  \(I^{\tau}_{\mathbb{S}_v} \in \mathbb{N}\).
The greater the SATB value of a satellite node, the higher its rank, indicating its more important role in the satellite network. 
Thus, \(I^{\tau}_{\mathbb{S}_v}\) provides a ranking for each critical node \(\mathbb{S}_v \in \mathcal{C}_{\tau} \), where a higher value corresponds to a satellite node whose failure would lead to a more severe disruption in networking functionality.

\begin{algorithm}[t]
\caption{Identification and ranking of critical nodes}
\label{alg:identification}
\begin{algorithmic}[1]
    \item \textbf{Input:}  The discrete temporal graphs \(\{\mathcal{G}_{\tau_1}, \mathcal{G}_{\tau_2}, \ldots, \mathcal{G}_{\tau_n}\}\), the set of active source-destination cell pairs \( \mathcal{H} = \{    (\mathbb{C}_i, \mathbb{C}_j) | \mathbb{C}_i, \mathbb{C}_j \in \mathcal{C}_{\text{active}}, \mathcal{I}_{ij} = 1 \}\), and satellite set \( \mathcal{S} \).
\item \textbf{Output:} The SATB for all satellites \( \mathcal{B} = \{ \mathcal{B}_{\mathbb{S}_v} |   \mathbb{S}_v \in \mathcal{S}\} \); the SATB matrix \(\mathbf{M}_{n \times |\mathcal{S}|}\); the ranking matrix \(\mathbf{P}_{n \times |\mathcal{S}|}\), where each row corresponds to a time window, and each column gives the rank order of satellites; and a dictionary $\mathcal{X} = \{(\mathbb{S}_v, \tau_x):  X^{\tau_x}_{\mathbb{S}_{v}}\}$, where the list \(X^{\tau_x}_{\mathbb{S}_{v}}\) stores the shortest paths that pass the node \(\mathbb{S}_v\) in time window $\tau_x$. 
    \item \textbf{Step 1: Calculate SATB based on \textbf{Algorithm \ref{computation-SATB-all}}.}
    \item Initialize the SATB vector \( \mathcal{B}_{\mathbb{S}_v} = \mathbf{0},  \forall \mathbb{S}_v \in \mathcal{S} \).
    \item Initialize \( X^{\tau_x}_{\mathbb{S}_{v}} = [\ ],  \forall \mathbb{S}_v \in \mathcal{S},  \forall 
 \tau_x \in \{\tau_1, \tau_2, \ldots, \tau_n\} \).
\item \textbf{for} each time window \( \tau_x \in \{\tau_1, \tau_2, \ldots, \tau_n\} \) \textbf{do}
 \item \ \ \  Initialize \( \mathcal{L} = \emptyset\) to store intermediate satellite nodes.
 \item \ \ \ \textbf{for} each source-destination cell pair \((\mathbb{C}_{i},\mathbb{C}_{j}) \in \mathcal{H} \)  \textbf{do}
       \item \ \ \ \ \ \ Find the shortest path \(p^{\tau_x}_{\mathbb{C}_{i} \rightarrow \mathbb{C}_{j}}\) in \(\mathcal{G}_{\tau_x}\).
 \item \ \ \ \ \ \ \textbf{for} each \(\mathbb{S}_\epsilon \in p^{\tau_x}_{\mathbb{C}_i \to \mathbb{C}_j}, v \notin \{\mathbb{C}_i, \mathbb{C}_j\} \)  \textbf{do}
   \item \ \ \ \ \ \  \ \ \  Update \(   \mathcal{L} = \mathcal{L} \cup \left\{ \mathbb{S}_\epsilon \right\}  \), \(  X^{\tau_x}_{\mathbb{S}_{v}} =X^{\tau_x}_{\mathbb{S}_{v}} \cup \left\{ p^{\tau_x}_{\mathbb{C}_i \to \mathbb{C}_j} \right\} \).
\item \ \ \  Let \( D = \{ \mathbb{S}_v :  \text{freq}( \mathbb{S}_v) = \text{count of }  \mathbb{S}_v \text{ in } \mathcal{L} \mid \mathbb{S}_v  \in \mathcal{L} \} \). 
\item \ \ \ \textbf{for}  each satellite node \( \mathbb{S}_v \) in \( D \) \textbf{do}
   \item \ \ \ \ \ \  Update \(\beta^{\tau_x}_{\mathbb{S}_v} = \text{freq}( \mathbb{S}_v) \). 
   \item \textbf{Step 2: Construct the SATB matrix \(\mathbf{M}_{n \times |\mathcal{S}|}\) using $\mathcal{B}$.}
    \item Construct \(\mathbf{M}_{n \times |\mathcal{S}|}\), where each row \(\mathbf{M}[i, :]\) represents the SATB values of all satellites in the time window \( \tau_i \).
\item \textbf{Step 3: Construct the ranking matrix \(\mathbf{P}_{n \times |\mathcal{S}|}\).}
    \item \textbf{for} each time window \( \tau_i \in \{\tau_1, \tau_2, \ldots, \tau_n\} \) \textbf{do}
    \item \ \ \ Define \(\pi_i\) as the descending sort permutation of \(\mathbf{M}[i, :]\), i.e., \(\mathbf{M}[i, \pi_i(1)] \geq \mathbf{M}[i, \pi_i(2)] \geq \dots \geq \mathbf{M}[i, \pi_i(|\mathcal{S}|)]\).
\item \ \ \ Set \(\mathbf{P}[i, j] = \pi_i(j)\) for \(j = 1, \dots, |\mathcal{S}|\), where \(\pi_i(j)\) gives the satellite ID of the \(j\)-th ranked SATB in \(\tau_i\).
    \item \textbf{return} \( \mathcal{B} \), $\mathcal{X}$,  \(\mathbf{M}_{n \times |\mathcal{S}|}\), and \(\mathbf{P}_{n \times |\mathcal{S}|}\).
\end{algorithmic}
\end{algorithm}

The identification and ranking procedure of critical nodes is outlined in \textbf{Algorithm \ref{alg:identification}}. The steps are as follows:
\begin{enumerate}
    \item \textbf{SATB calculation}: For each time window \(\tau_x\), we calculate the SATB value \(\beta^{\tau_x}_{\mathbb{S}_v}\) for each satellite \(\mathbb{S}_v\) based on \textbf{Algorithm \ref{computation-SATB-all}}. We iterate over each active source-destination pair, computing shortest paths and incrementing the SATB of the nodes on these paths. The paths traversed by each satellite node at each time window \(\tau_x\) are recorded  in a dictionary \( \mathcal{X} = \{(\mathbb{S}_v, \tau_x): X^{\tau_x}_{\mathbb{S}_v}\} \), which is essential for evaluating the impact of node failures. 
    
    \item \textbf{SATB matrix construction}: The SATB values are aggregated across time windows into a matrix \(\mathbf{M}_{n \times |\mathcal{S}|}\), where each row \(\mathbf{M}[i, :]\) corresponds to the SATB values for all satellites in time window \(\tau_i\). This matrix gives a temporal overview of each node's importance, assisting in assessing the overall impact of each satellite.
    
    \item \textbf{Ranking of nodes by importance}: To rank the nodes, we sort each row of the SATB matrix \(\mathbf{M}_{n \times |\mathcal{S}|}\) in descending order, producing a ranking matrix \(\mathbf{P}_{n \times |\mathcal{S}|}\). The element \(\mathbf{P}[i, j]\) represents the rank of satellite \(\mathbb{S}_j\) in time window \(\tau_i\) based on its SATB value. This ranking helps track each node's importance and identify which nodes are more critical to maintaining network service.
\end{enumerate}

Based on \textbf{Algorithm \ref{alg:identification}}, the ranking indicators \( I^{\tau}_{\mathbb{S}_v} \) for each satellite node \(\mathbb{S}_v\) at time window \(\tau\) can be determined directly from the ranking matrix \(\mathbf{P}_{n \times |\mathcal{S}|}\). The matrix \( \mathbf{P} \) is constructed by sorting the SATB values for each time window and assigning ranks accordingly. The position of a node \(\mathbb{S}_v\) in the matrix \( \mathbf{P} \) at a given time window \(\tau\) corresponds to the rank \( I^{\tau}_{\mathbb{S}_v} \), which indicates its relative importance. The nodes with the highest ranks in the matrix correspond to the most critical nodes for inter-satellite networking during that time window.

As a result, the set of critical nodes \(\mathcal{C}_{\tau}\) at time window \(\tau\)  can be determined by extracting the non-zero elements from the \( \tau \)-th row of the ranking matrix \(\mathbf{P}_{n \times |\mathcal{S}|}\). These elements represent the nodes that are actively involved in facilitating networking during the time window \(\tau\).
Thus, the ranking matrix \(\mathbf{P}_{n \times |\mathcal{S}|}\) not only helps determine the rank \( I^{\tau}_{\mathbb{S}_v} \) of each node, but also facilitates the identification of critical nodes \(\mathcal{C}_{\tau}\) by simply extracting the non-zero rankings from the corresponding row.

\vspace{-3mm}

\subsection{Impact Analysis of Node Failures on Inter-satellite Networking}

In this subsection, we analyze how node failures affect inter-satellite networking between ground cells. When a node \( \mathbb{S}_f \) fails and is critical in a time window $\tau_x$, (i.e., \( | X^{\tau_x}_{\mathbb{S}_f} | > 0 \)), all paths passing through  \( \mathbb{S}_f  \)  will be disrupted. Consequently, it will be necessary to recalculate all the affected paths in the list  \( X^{\tau_x}_{\mathbb{S}_f}\). This recalculation not only alters the routing path itself but also impacts the importance of all nodes along both the original paths and the newly computed ones. 
Furthermore, changes in the routing path introduce variations in the end-to-end delay. 
In the most extreme case, the failure of a critical node may render no alternate path to exist, compromising the service connectivity between the involved ground cells.
Thus, the failure of nodes can have a profound effect on various aspects of inter-satellite networking. 
We focus on the following key consequences of node failures:
\begin{itemize}
    \item \textbf{Impact on service connectivity}: The node failure event may reduce the number of reachable cell pairs, thus diminishing overall service connectivity across the network.

    \item \textbf{Impact on path delay}: Changes in end-to-end propagation delay due to affected path recalculations.

       \item \textbf{Impact on other node's importance}: The node failure event may alter the importance of other satellite nodes as their roles in affected paths change over time.
    
\end{itemize}

To quantify these impacts, for each active cell pair \( (\mathbb{C}_s, \mathbb{C}_d)\), we define the end-to-end propagation delay \( L_{{p}^{\tau_x}_{\mathbb{C}_s \rightarrow \mathbb{C}_d}} \) for the shortest path ${p}^{\tau_x}_{\mathbb{C}_s \rightarrow \mathbb{C}_d}$ during time window \( \tau_x \) as:
\begin{equation}
    L_{{p}^{\tau_x}_{\mathbb{C}_s \rightarrow \mathbb{C}_d}} = \sum_{(u, v) \in {p}^{\tau_x}_{\mathbb{C}_s \rightarrow \mathbb{C}_d}} L^{\tau_x}_{u, v},
\end{equation}
where \( L^{\tau_x}_{u, v} \) is the propagation delay of link \( (u, v) \) during \( \tau_x \).

In addition, we define the \textit{service connectivity ratio} \( \eta_{\tau_x} \) in time window \( \tau_x \) to assess the service connectivity within the satellite network, which is given by:
\begin{equation}
    \eta_{\tau_x} = \frac{|Q_{\tau_x}|}{|\mathcal{H}|} \times 100\%,
\end{equation}
where \( |\mathcal{H}| = |\{   (\mathbb{C}_i, \mathbb{C}_j) | \mathbb{C}_i, \mathbb{C}_j \in \mathcal{C}_{\text{active}}, \mathcal{I}_{ij} = 1 \}| \) is the total number of active ground cell pairs, and \( Q_{\tau_x} = \{ q^{\tau_x}_{\mathbb{C}_s, \mathbb{C}_d} | (\mathbb{C}_s, \mathbb{C}_d)  \in \mathcal{H}\} \) is the set of connected ground cell pairs in \( \tau_x \in \{\tau_1, \tau_2, \ldots, \tau_n\} \). For each active cell pair \( (\mathbb{C}_s, \mathbb{C}_d) \), the indicator variable \( q^{\tau_x}_{\mathbb{C}_s, \mathbb{C}_d} \in Q_{\tau_x} \) is defined as:
\begin{equation}
    q^{\tau_x}_{\mathbb{C}_s, \mathbb{C}_d} = 
\begin{cases} 
1, & \text{if }  L_{{p}^{\tau_x}_{\mathbb{C}_s \rightarrow \mathbb{C}_d}} \neq \infty, \\
0, & \text{if }  L_{{p}^{\tau_x}_{\mathbb{C}_s \rightarrow \mathbb{C}_d}} = \infty.
\end{cases}
\end{equation}
Here, \( q^{\tau_x}_{\mathbb{C}_s, \mathbb{C}_d} = 1 \) indicates that there is a valid path between ground cells \( \mathbb{C}_s \) and \( \mathbb{C}_d \) in time window $\tau_x$, as \( L_{{p}^{\tau_x}_{\mathbb{C}_s \rightarrow \mathbb{C}_d}} \neq \infty \); otherwise, \( q^{\tau_x}_{\mathbb{C}_s, \mathbb{C}_d} = 0 \), indicating no available path exists.
Therefore, a drop in  \( \eta_{\tau_x} \) indicates that some active ground cell pairs have lost connectivity due to node failure event.

In the following, we propose the node failure impact evaluation algorithms that takes as input the node failure event to quantitatively assess its impact on service connectivity, path delay, and the importance of other operational satellites.

\subsubsection{Algorithm of impact evaluation for single node failure}

\begin{algorithm}[t]
\caption{Impact evaluation for single node failure}
\label{alg:single_node_failure}
\begin{algorithmic}[1]
    \item \textbf{Input:} The discrete temporal graphs \(\{\mathcal{G}_{\tau_1}, \mathcal{G}_{\tau_2}, \ldots, \mathcal{G}_{\tau_n}\}\), the node failure event  \( F = (t_f, \mathcal{S}_f = \{ \mathbb{S}_f\}), t_f \in \tau_f \).
    \item \textbf{Output:} Updated ratio of reachable cell pairs  \(\eta = \{ \eta_{\tau_x} |   \tau_x \in \{\tau_1, \tau_2, \ldots, \tau_n\} \} \), the affected paths $\mathcal{P}_{\mathbb{S}_{f}}$, and a set of $\mathbb{S}_{f}$-affected satellite nodes \(W_{\mathbb{S}_{f}}\).
    \item \textbf{for} each time window \({\tau_x} \geq \tau_f \) \textbf{do}
        \item \ \ \ \ Remove the failed node \(\mathbb{S}_{f}\) from \(\mathcal{G}_{\tau_x}\) to build \(\mathcal{G'}_{\tau_x}\).
    \item \ \ \ \ Obtain $\mathbb{S}_{f}$-affected path list \( X^{\tau_x}_{\mathbb{S}_{f}} \) at time window \(\tau_x\).
    \item \ \ \ \ \textbf{if} $ |X^{\tau_x}_{\mathbb{S}_{f}}| = 0$ \textbf{then}
    \item \ \ \ \ \ \ \ \  \textbf{continue} // $\mathbb{S}_{f}$ is not a critical node during $\tau_x$. 
    \item \ \ \ \ \textbf{for} each path ${p}^{\tau_x}_{\mathbb{C}_s \rightarrow \mathbb{C}_d}$  in \( X^{\tau_x}_{\mathbb{S}_{f}} \) \textbf{do}
   \item \ \ \ \ \ \ \ \ 
 Update  $\mathcal{P}_{\mathbb{S}_{f}} = \mathcal{P}_{\mathbb{S}_{f}} \cup \{ {p}^{\tau_x}_{\mathbb{C}_s \rightarrow \mathbb{C}_d}: (L_{{p}^{\tau_x}_{\mathbb{C}_s \rightarrow \mathbb{C}_d}}, + \infty)\}$.
    \item \ \ \ \ \ \ \ \ \textbf{for} each satellite node $\mathbb{S}_{y}$ on the path ${p}^{\tau_x}_{\mathbb{C}_s \rightarrow \mathbb{C}_d}$ \textbf{do}
    \item \ \ \ \ \ \ \ \ \ \ \ \ Update $\beta^{\tau_x}_{\mathbb{S}_{y}}  = \beta^{\tau_x}_{\mathbb{S}_{y}} -1$.
    \item \ \ \ \ \ \ \ \ \ \ \ \ Remove the path ${p}^{\tau_x}_{\mathbb{C}_s \rightarrow \mathbb{C}_d}$ from \( X^{\tau_x}_{\mathbb{S}_{y}} \).
    \item \ \ \ \ \ \ \ \ \ \ \ \  Update the set \(W_{\mathbb{S}_{f}} 
 = W_{\mathbb{S}_{f}} \cup \{\mathbb{S}_{y}\}\).
    \item \ \ \ \ \ \ \ \  Recalculate the shortest path ${p'}^{\tau_x}_{\mathbb{C}_s \rightarrow \mathbb{C}_d}$ in $\mathcal{G'}_{\tau_x}$.
    \item \ \ \ \ \ \ \ \ \textbf{if} a shortest path ${p'}^{\tau_x}_{\mathbb{C}_s \rightarrow \mathbb{C}_d}$ exists in $\mathcal{G'}_{\tau_x}$ \textbf{then}
    \item \ \ \ \ \ \ \ \ \ \ \ \  Update  $\mathcal{P}_{\mathbb{S}_{f}}( {p}^{\tau_x}_{\mathbb{C}_s \rightarrow \mathbb{C}_d})=(L_{{p}^{\tau_x}_{\mathbb{C}_s \rightarrow \mathbb{C}_d}}, L_{{p'}^{\tau_x}_{\mathbb{C}_s \rightarrow \mathbb{C}_d}})\}$
    \item \ \ \ \ \ \ \ \ \ \ \ \  \textbf{for} each satellite node $\mathbb{S}_{z}$ on ${p'}^{\tau_x}_{\mathbb{C}_s \rightarrow \mathbb{C}_d}$ \textbf{do}
    \item \ \ \ \ \ \ \ \ \ \ \ \ \ \ \ \  Update $\beta^{\tau_x}_{\mathbb{S}_{z}}  = \beta^{\tau_x}_{\mathbb{S}_{z}} +1$.
    \item \ \ \ \ \ \ \ \ \ \ \ \ \ \ \ \ Add the new shortest path ${p'}^{\tau_x}_{\mathbb{C}_s \rightarrow \mathbb{C}_d}$ \( X^{\tau_x}_{\mathbb{S}_{z}} \).
    \item \ \ \ \ \ \ \ \ \  \ \ \ \ \ \ \ Update the set \(W_{\mathbb{S}_{f}} 
 = W_{\mathbb{S}_{f}} \cup \{\mathbb{S}_{z}\}\).
    \item \ \ \ \ \ \ \ \ \textbf{else} // The post-failure shortest path does not exist. 
    \item \ \ \ \ \ \ \ \ \ \ \ \ Update the ratio of reachable cell pairs $\eta_{\tau_x}$.
    \item \ \ \ \ \ \ \ \ \ \ \ \ Set the end-to-end delay of the path to infinity.
    \item \textbf{return}  \(\eta = \{ \eta_{\tau_x} |   \tau_x \in \{\tau_1, \tau_2, \ldots, \tau_n\} \} \),  $\mathcal{P}_{\mathbb{S}_{f}}$, and \(W_{\mathbb{S}_{f}}\).
\end{algorithmic}
\end{algorithm}

Given a single node failure event \( F = (t_f, \mathcal{S}_f = \{ \mathbb{S}_f\}), t_f \in \tau_f \), we evaluate its impact based on the discrete temporal graphs \(\{\mathcal{G}_{\tau_1}, \mathcal{G}_{\tau_2}, \ldots, \mathcal{G}_{\tau_n}\}\). The evaluation procedure is outlined in \textbf{Algorithm \ref{alg:single_node_failure}}. First, for each time window \(\tau_x \geq \tau_f\), we construct an updated temporal graph \(\mathcal{G'}_{\tau_x}\) by removing the failed node \(\mathbb{S}_f\) from \(\mathcal{G}_{\tau_x}\).
Next, we examine all end-to-end shortest paths in \(\mathcal{G}_{\tau_x}\) that originally traversed the failed node \(\mathbb{S}_f\), recording them in the affected path set \(X^{\tau_x}_{\mathbb{S}_{f}}\). For each path \({p}^{\tau_x}_{\mathbb{C}_s \rightarrow \mathbb{C}_d} \in X^{\tau_x}_{\mathbb{S}_{f}}\), we record the original path delay and then set the path delay as infinite, marking it as disrupted in the affected path record using a dictionary \(\mathcal{P}_{\mathbb{S}_{f}}\). Additionally, we adjust the SATB value \(\beta^{\tau_x}_{\mathbb{S}_{y}}\) of each satellite \(\mathbb{S}_{y}\) on these affected paths and add \(\mathbb{S}_{y}\) into the set of affected nodes \(W_{\mathbb{S}_{f}}\).

Following this, we attempt to identify a new shortest path \({p'}^{\tau_x}_{\mathbb{C}_s \rightarrow \mathbb{C}_d}\) in the updated temporal graph \(\mathcal{G'}_{\tau_x}\). If a new path is found, we update the dictionary \(\mathcal{P}_{\mathbb{S}_{f}}\) with both the original and recalculated path delays, adjust the SATB value for each node on the new path, and add these affected nodes on the new path \({p'}^{\tau_x}_{\mathbb{C}_s \rightarrow \mathbb{C}_d}\) to \(W_{\mathbb{S}_{f}}\). If no alternative path exists, we update the service connectivity ratio \(\eta_{\tau_x}\) to reflect the reduction in reachable cell pairs, noting an infinite delay for the path.
Finally, the output includes the set \(\eta\) of updated service connectivity ratios across time windows, the record of affected paths \(\mathcal{P}_{\mathbb{S}_{f}}\) with delay changes, and the set of affected other satellite nodes \(W_{\mathbb{S}_{f}}\).

\textbf{Complexity Analysis:}  
The complexity of \textbf{Algorithm \ref{alg:single_node_failure}} depends on the number of time windows and affected paths. For each time window \(\tau_x\), updating \(\mathcal{G'}_{\tau_x}\) has \(O(1)\) complexity, and recalculating paths in \(X^{\tau_x}_{\mathbb{S}_f}\) requires \(O(|X^{\tau_x}_{\mathbb{S}_f}| \cdot |\mathcal{V}| \log |\mathcal{V}|)\) operations. Thus, the overall complexity is \(O(n \cdot |X^{\tau_x}_{\mathbb{S}_f}| \cdot |\mathcal{V}| \log |\mathcal{V}|)\).

\subsubsection{Algorithm of impact evaluation for multiple node Failure}

\begin{algorithm}[t]
\caption{Impact evaluation for multiple node failures}
\label{alg:multiple_node_failures}
\begin{algorithmic}[1]
    \item \textbf{Input:} The discrete temporal graphs \(\{\mathcal{G}_{\tau_1}, \mathcal{G}_{\tau_2}, \ldots, \mathcal{G}_{\tau_n}\}\), the node failure event  \( F = (t_f, \mathcal{S}_f), |\mathcal{S}_f|>1, t_f \in \tau_f \).
    \item \textbf{Output:} Updated ratio of reachable cell pairs  \(\eta = \{ \eta_{\tau_x} |   \tau_x \in \{\tau_1, \tau_2, \ldots, \tau_n\} \} \), the affected paths \(\mathcal{P}_{\mathcal{S}_{f}}\), and a set of affected satellite nodes \(W_{\mathcal{S}_{f}}\).
    \item \textbf{for} each time window \({\tau_x} \geq \tau_f \) \textbf{do}
       \item \ \ \ \  Initialize the full set of the affected paths $X_{F} = \emptyset$. 
        \item \ \ \ \ Remove the failed nodes \(\mathcal{S}_{f}\) from \(\mathcal{G}_{\tau_x}\) to build \(\mathcal{G'}_{\tau_x}\).
    \item \ \ \ \ \textbf{for} each failed node \(\mathbb{S}_f \in \mathcal{S}_{f}\) \textbf{do}
    \item \ \ \ \ \ \ \ \  Obtain \(\mathbb{S}_f\)-affected path list \( X^{\tau_x}_{\mathbb{S}_f} \) at \(\tau_x\).
        \item \ \ \ \ \ \ \ \ Update $X_{F}  = X_{F}  \cup X^{\tau_x}_{\mathbb{S}_f}$. 
    \item \ \ \ \  \textbf{for} each path \({p}^{\tau_x}_{\mathbb{C}_s \rightarrow \mathbb{C}_d}\) in \( X_{F} \) \textbf{do}
    \item \ \ \ \ \ \ \ \ Update \(\mathcal{P}_{\mathcal{S}_{f}}[{p}^{\tau_x}_{\mathbb{C}_s \rightarrow \mathbb{C}_d}] = (L_{{p}^{\tau_x}_{\mathbb{C}_s \rightarrow \mathbb{C}_d}}, +\infty)\).
    \item \ \ \ \ \ \ \ \ \textbf{for} each satellite \(\mathbb{S}_y\) on the path \({p}^{\tau_x}_{\mathbb{C}_s \rightarrow \mathbb{C}_d}\) \textbf{do}
    \item \ \ \ \ \ \ \ \ \ \ \ \ Update \(\beta^{\tau_x}_{\mathbb{S}_y} = \beta^{\tau_x}_{\mathbb{S}_y} - 1\).
    \item \ \ \ \ \ \ \ \ \ \ \ \ Remove the path \({p}^{\tau_x}_{\mathbb{C}_s \rightarrow \mathbb{C}_d}\) from \( X^{\tau_x}_{\mathbb{S}_y} \).
    \item \ \ \ \ \ \ \ \ \ \ \ \ Update the set \( W_{\mathcal{S}_{f}} = W_{\mathcal{S}_{f}} \cup \{\mathbb{S}_y\} \).
    \item \ \ \ \  \ \ \ \ Recalculate the shortest path \({p'}^{\tau_x}_{\mathbb{C}_s \rightarrow \mathbb{C}_d}\) in \(\mathcal{G'}_{\tau_x}\).
    \item \ \ \ \ \ \ \ \  \textbf{if} a shortest path \({p'}^{\tau_x}_{\mathbb{C}_s \rightarrow \mathbb{C}_d}\) exists in \(\mathcal{G'}_{\tau_x}\) \textbf{then}
    \item \ \ \ \ \ \ \ \  \ \ \ \ Update \(\mathcal{P}_{\mathcal{S}_{f}}[{p}^{\tau_x}_{\mathbb{C}_s \rightarrow \mathbb{C}_d}] = (L_{{p}^{\tau_x}_{\mathbb{C}_s \rightarrow \mathbb{C}_d}}, L_{{p'}^{\tau_x}_{\mathbb{C}_s \rightarrow \mathbb{C}_d}})\).
    \item \ \ \ \ \ \ \ \  \ \ \ \ \textbf{for} each satellite node \(\mathbb{S}_z\) on \({p'}^{\tau_x}_{\mathbb{C}_s \rightarrow \mathbb{C}_d}\) \textbf{do}
    \item \ \ \ \ \ \ \ \ \ \ \ \  \ \ \ \ Update \(\beta^{\tau_x}_{\mathbb{S}_z} = \beta^{\tau_x}_{\mathbb{S}_z} + 1\).
    \item \ \ \ \ \ \ \ \ \ \ \ \  \ \ \ \ Add the new path to the node's list \( X^{\tau_x}_{\mathbb{S}_z} \).
    \item \ \ \ \ \ \ \ \ \ \ \ \ \ \ \ \  Update the set \( W_{\mathcal{S}_{f}} = W_{\mathcal{S}_{f}} \cup \{\mathbb{S}_z\} \).
    \item \ \ \ \   \ \ \ \ \textbf{else} // The post-failure shortest path does not exist
    \item \ \ \ \ \ \ \ \  \ \ \ \ Update the ratio of reachable cell pairs \(\eta_{\tau_x}\).
    \item \ \ \ \ \ \ \ \   \ \ \ \ Set the end-to-end delay of the path to infinity.
    \item \textbf{return} \(\eta = \{ \eta_{\tau_x} |   \tau_x \in \{\tau_1, \tau_2, \ldots, \tau_n\} \} \), \(\mathcal{P}_{\mathcal{S}_{f}}\), and \( W_{\mathcal{S}_{f}} \).
\end{algorithmic}
\end{algorithm}

To assess the impact of multiple node failures, we propose an efficient approach that builds upon and extends the method for single node failure evaluation. Given a  multiple node failure event \( F = (t_f, \mathcal{S}_f), |\mathcal{S}_f| > 1, t_f \in \tau_f \), \textbf{Algorithm \ref{alg:multiple_node_failures}} iteratively evaluates its impact across temporal graphs \(\{\mathcal{G}_{\tau_1}, \mathcal{G}_{\tau_2}, \ldots, \mathcal{G}_{\tau_n}\}\). 
First, in each time window \(\tau_x \geq \tau_f\), we update the temporal graph by removing all failed nodes in \( \mathcal{S}_f \), resulting a modified graph \(\mathcal{G'}_{\tau_x}\). Unlike the single node failure evaluation, which processes paths separately for each node, the multiple node failure algorithm aggregates paths affected by all failed nodes into a unified set \( X_F \). This path aggregation simplifies the process by consolidating overlapping paths that may be affected by multiple failed nodes, reducing computational redundancy and enhancing efficiency.

For each affected path in \( X_F \), we update the affected paths record \(\mathcal{P}_{\mathcal{S}_f}\) to reflect the pre-failure and post-failure status, setting the affected paths' delays to infinity and adjusting the SATB value \(\beta^{\tau_x}_{\mathbb{S}}\) of each involved satellite node. If alternative shortest paths exist in \(\mathcal{G'}_{\tau_x}\), these paths are updated and the associated records are updated. If no alternative path is found, the service connectivity ratio \(\eta_{\tau_x}\) is updated and reduced accordingly. 
By centering the analysis on aggregated affected paths by multiple node failure, \textbf{Algorithm \ref{alg:multiple_node_failures}} optimizes evaluation procedure and provides a scalable framework for assessing multi-node failure scenarios in mega-satellite constellations.

\textbf{Complexity Analysis:}  
The complexity of \textbf{Algorithm \ref{alg:multiple_node_failures}} depends on the number of failed nodes and time windows. Updating a temporal graph \(\mathcal{G'}_{\tau_x}\) requires \(O(|\mathcal{S}_f|)\) operations. Recalculating paths requires \(O(|X_F| \cdot |\mathcal{V}| \log |\mathcal{V}|)\) operations. Thus, the overall complexity is \(O(n \cdot (|\mathcal{S}_f| + |X_F| \cdot |\mathcal{V}| \log |\mathcal{V}|))\).

\vspace{-3 mm}
\section{Performance Evaluation}  \label{sec:evaluation}

\begin{figure*}[htbp]  % Use figure* to span both columns
    \centering
    \begin{minipage}[t]{0.45\textwidth}
        \centering
        \includegraphics[width=60mm]{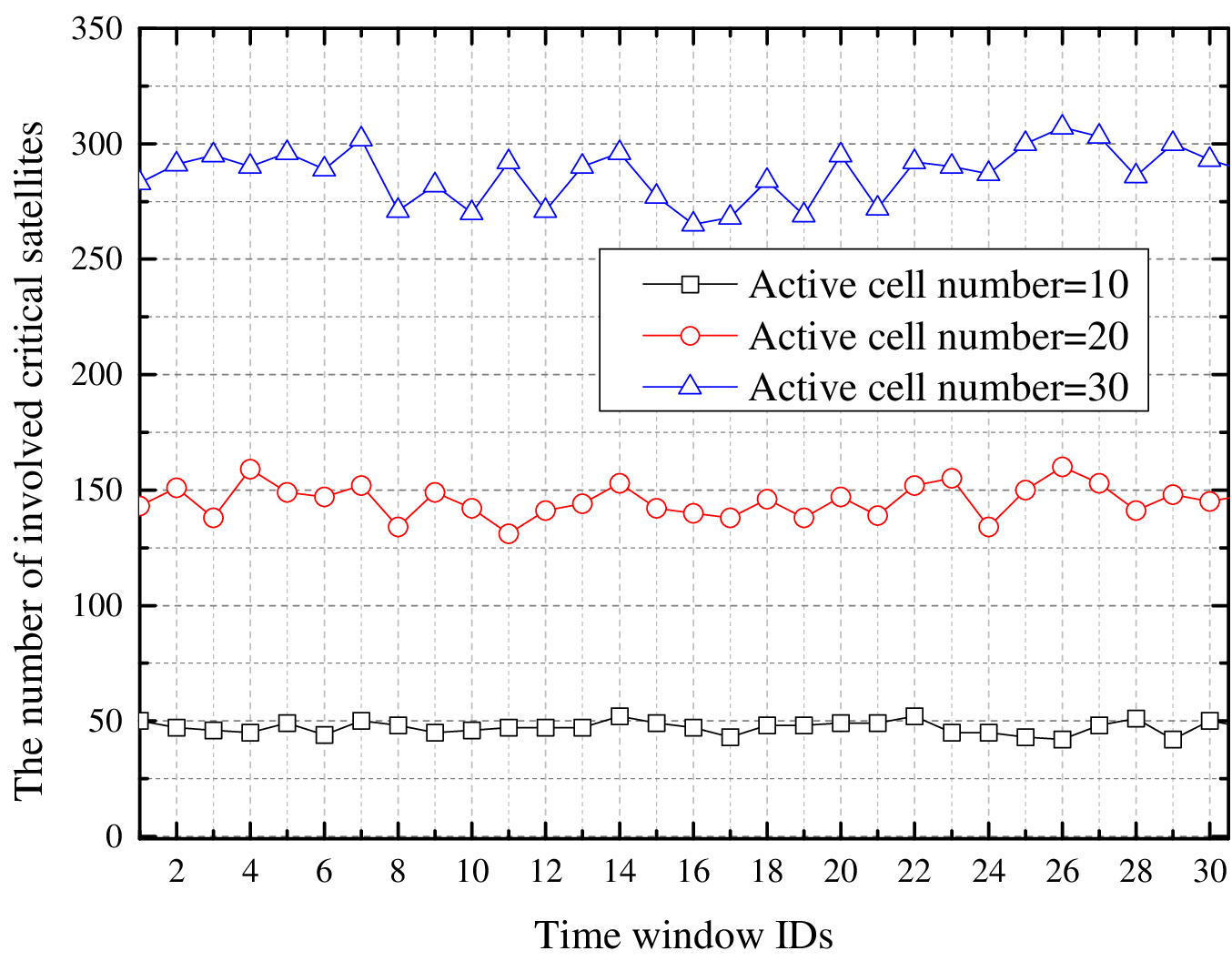}
        \caption{The number of critical satellites under normal conditions.}
        \label{fig:num_of_critical_satellite}
    \end{minipage}%
    \hspace{0.05\textwidth}  % Adjust space between the figures
    \begin{minipage}[t]{0.45\textwidth}
        \centering
                \includegraphics[width=60mm]{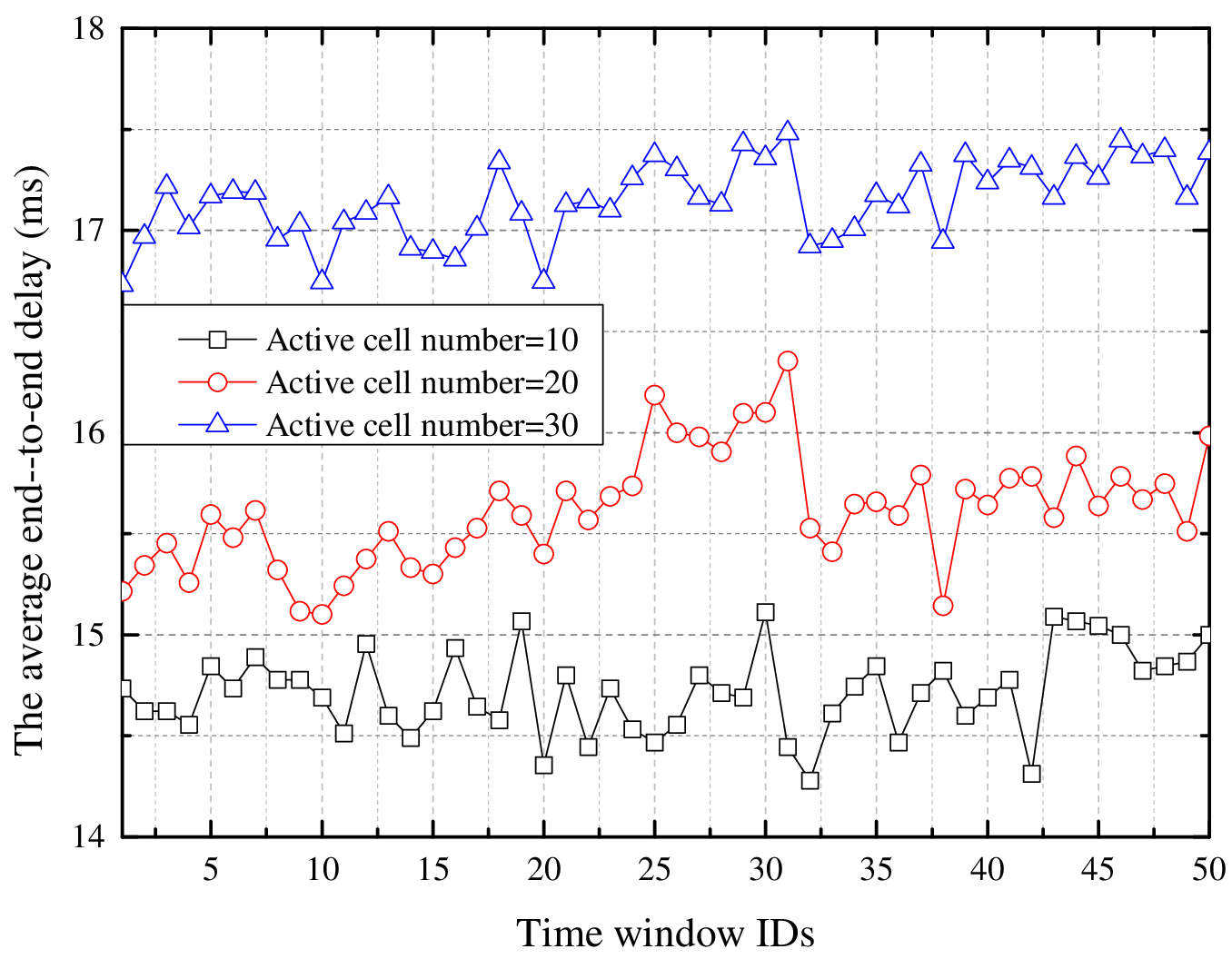}
        \caption{The average end-to-end path delay under normal conditions.}
        \label{fig:path_delay_over_time}
    \end{minipage}
     \vspace{- 3mm}
\end{figure*}

\subsection{Simulation Setup}
We conduct simulations using the Starlink constellation, where 2,000 Starlink satellites are selected from the standard object database provided by Systems Tool Kit (STK). Specifically, we select satellites from both the first and second deployment phases of the Starlink constellation. The first phase includes 1,600 satellites in 1,150 km circular orbits with an inclination of 53°, distributed across 32 orbital planes with 50 satellites per plane. The second phase contains a total of 2,825 satellites, from which we select 400 satellites in 1,130 km circular orbits with an inclination of 74°, organized in 8 orbital planes with 50 satellites per plane. These satellites thus belong to different orbital layers. 
Ground cells are distributed across 30 cities worldwide, including 
Xi'an, Beijing, Sanya, Kunming, Miyun, 
Kashi, Amsterdam, Athens, Barcelona, Berlin, 
Dubai, Istanbul, Kuala Lumpur, London, Madrid, 
Milan, Moscow, Mumbai, New York, Paris, 
Rio de Janeiro, Rome, Seoul, Shanghai, Singapore, 
Sydney, Tokyo, Toronto, Vienna, and Washington. Using the \textit{compute access} function in STK, we generate contact plans for SGLs and ISLs. SGLs and ISLs are determined by line-of-sight visibility: a communication opportunity is considered valid if the distance between a satellite and a ground cell, or between two satellites, is within 2,500~km. All communication opportunities within the visibility threshold are included in the link set for each time window as candidate transmission links for dynamic routing \cite{yao2025leo}. 
To model the time-varying nature of the network, we employ the enhanced temporal graph construction model described in  \cite{guo2024enhanced}, where the service duration threshold for time division is set as $60$ seconds. The propagation delays for both SGLs and ISLs are uniformly distributed between $5$ and $15$ ms as in \cite{guo2023online, guo2024lightweight}. 
The simulation period spans from 2024-10-09 04:00:00 to 2024-10-10 04:00:00, covering a full day and divided into multiple time windows. Each time window lasts at least 60 seconds, as the service duration threshold for time division is set to 60 seconds as previously noted. Each service involves a pair of ground cells with persistent communication demand throughout the entire simulation period, and routing paths are computed in every window to support continuous service over the whole day.
All algorithms are implemented in Python. 
For clarity, the simulation parameters are outlined in Table \ref{table:3}.

\begin{table}[h]
	\caption{Simulation parameters}
	 \vspace{-3 mm}
	\label{table:3}
	\begin{center}
		\begin{tabular}{|l|l|} 
			\hline
			\textbf{Parameter} & \textbf{Value} \\
			\hline
			simulation duration & 24 hours \\
			\hline
			number of satellites & 2000 Starlink satellites (1600+400) \\
			\hline
			orbital layers & 2 \\
			\hline
			orbital planes & initial (32), second (8) \\
			\hline
			sats per plane & initial (50), second (50) \\
			\hline
			altitude (km) & initial (1150), second (1130) \\
			\hline
			inclination (deg) & initial (53°), second (74°) \\
			\hline
			antenna type & steerable, radio frequency/laser-capable \\
			\hline
			connectivity model & dynamic, Line-of-Sight-based \\
            \hline
			number of ground cells & 30  \\
			\hline
			routing algorithm & delay-shortest path \\
			\hline
			propagation delay & 5–15 ms \\
			\hline
			Line-of-Sight threshold & 2500 km \\
			\hline
		\end{tabular}
	\end{center}
\end{table}

\vspace{-3mm}

\subsection{Simulation Results and Analysis}

\subsubsection{Evaluation of inter-satellite networking performance under normal conditions}

\begin{figure*}[htbp]  % Use figure* to span both columns
    \centering
    \begin{minipage}[t]{0.45\textwidth}
        \centering
              \includegraphics[width=80mm]{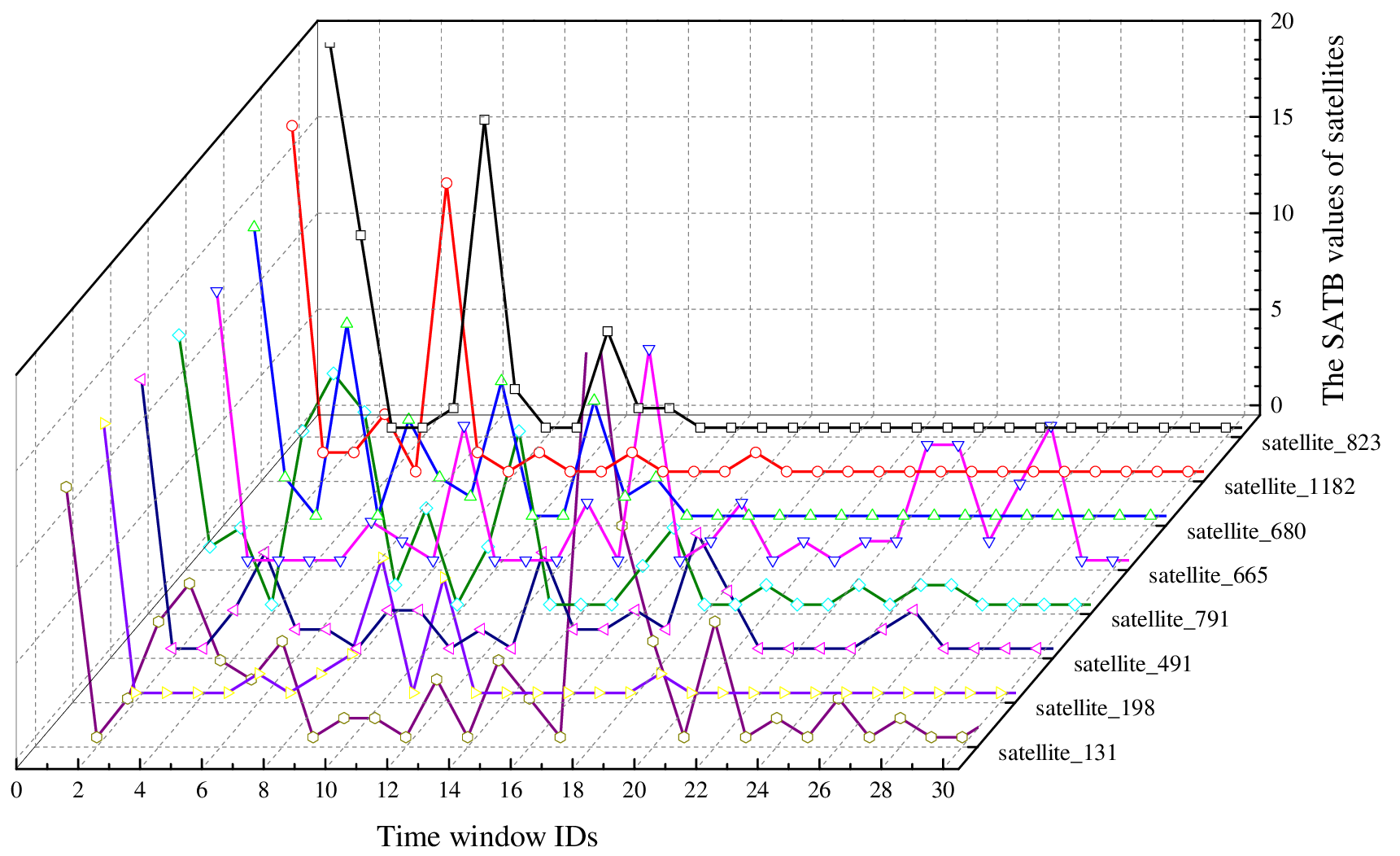}
        \caption{The SATB values of critical satellite nodes under normal conditions.}
\label{fig:SATB_over_time}
    \end{minipage}%
    \hspace{0.05\textwidth}  % Adjust space between the figures
    \begin{minipage}[t]{0.45\textwidth}
        \centering
          \includegraphics[width=80mm]{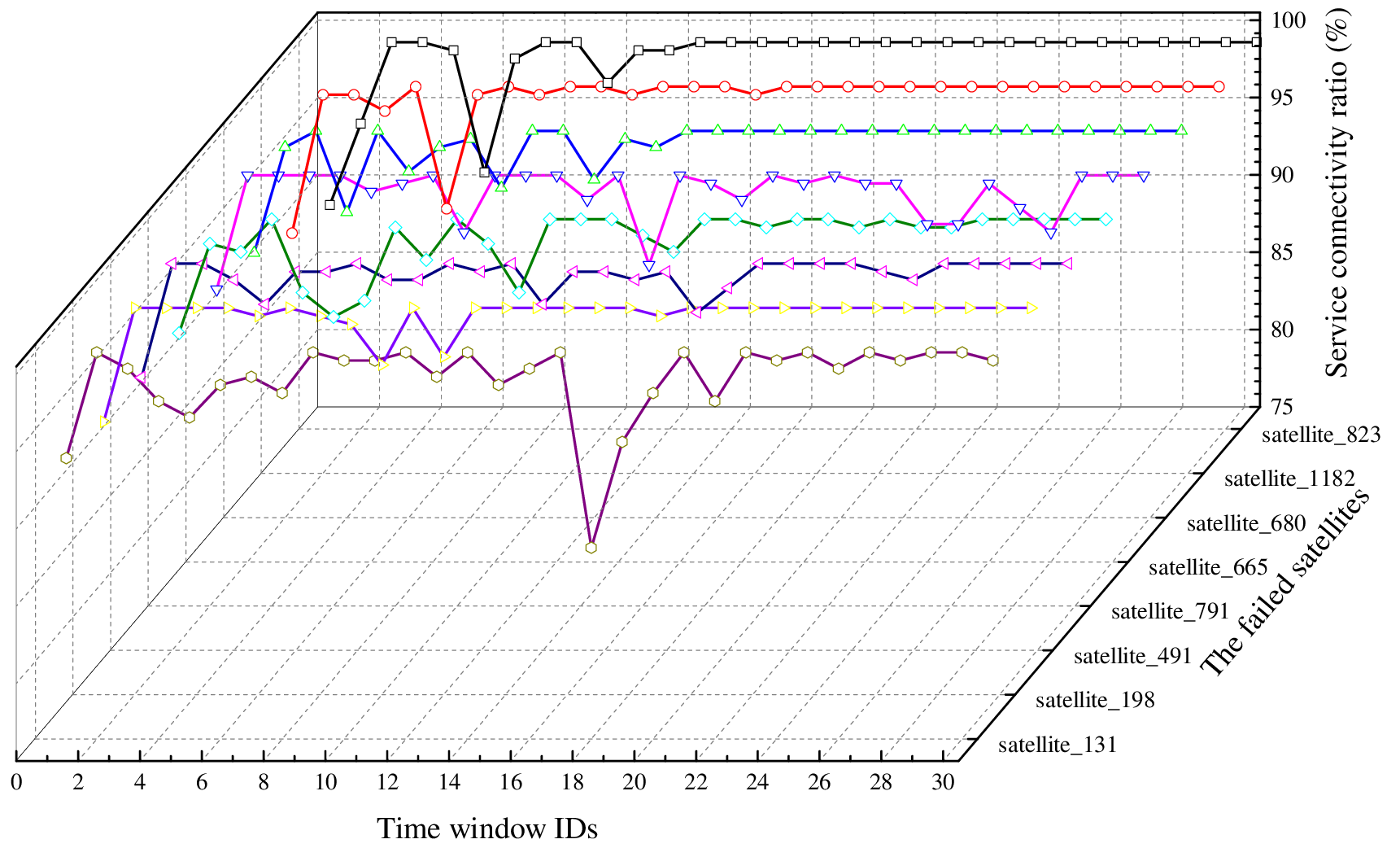}
        \caption{The service connectivity ratios under single node failure.}
\label{fig:ratio_single_node_failure}
    \end{minipage}
     \vspace{- 3mm}
\end{figure*}

\begin{figure*}[htbp]  % Use figure* to span both columns
    \centering
    \begin{minipage}[t]{0.45\textwidth}
        \centering
                \includegraphics[width=80mm]{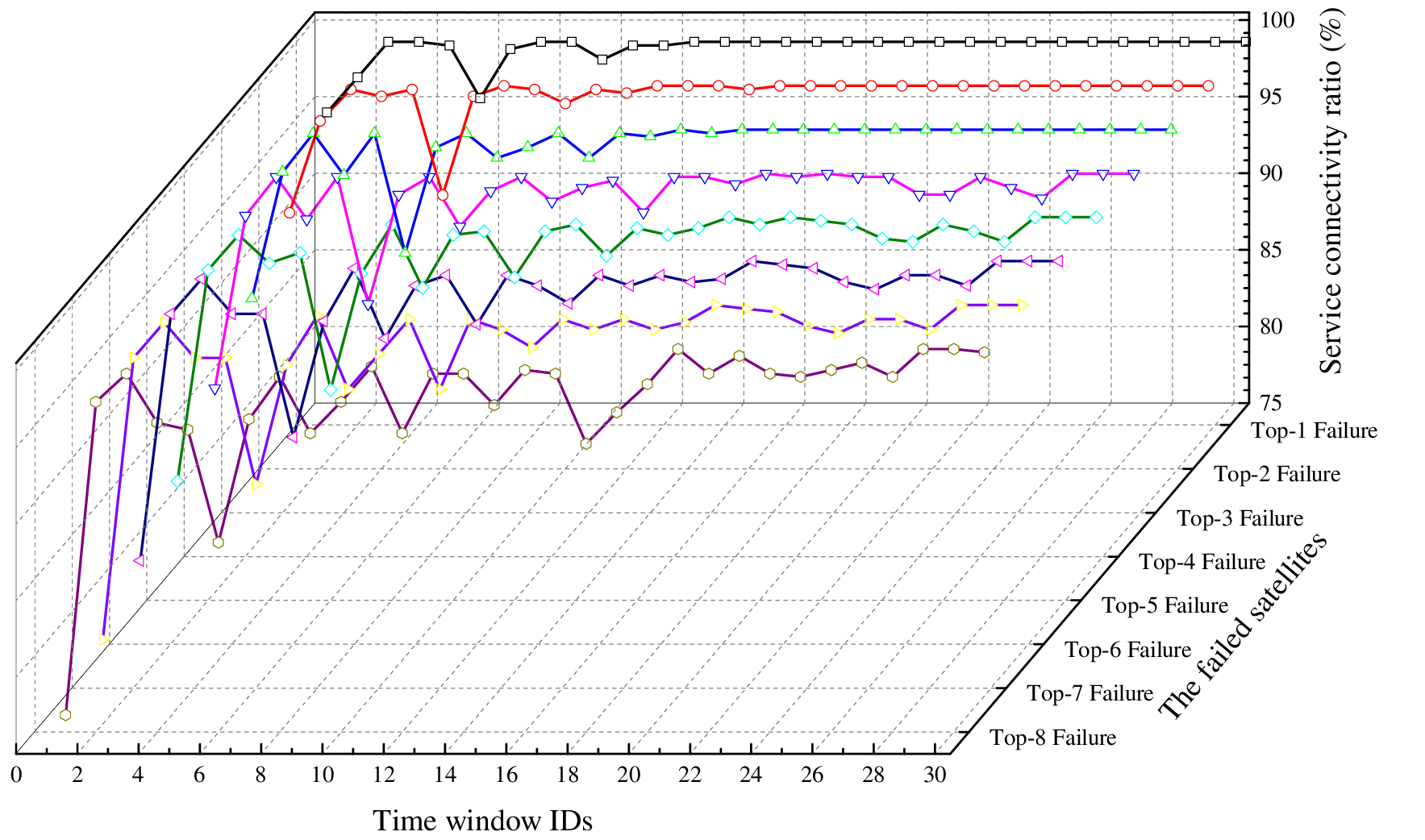}
       \caption{The service connectivity ratios under multiple node failure.}
\label{fig:ratio_multiple_node_failure}
    \end{minipage}%
    \hspace{0.05\textwidth}  % Adjust space between the figures
    \begin{minipage}[t]{0.45\textwidth}
        \centering
              \includegraphics[width=80mm]{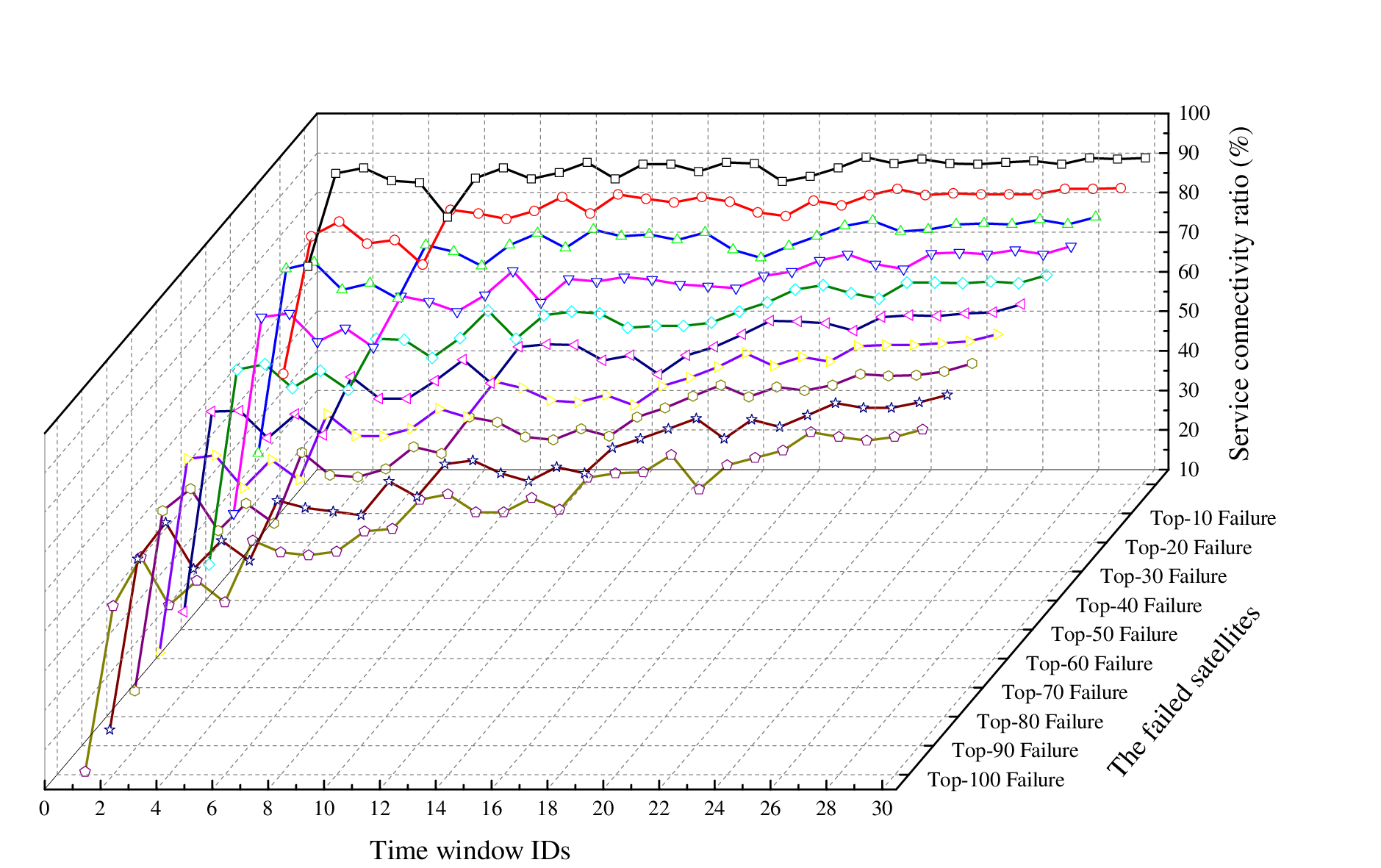}
       \caption{The service connectivity ratios under large-scale node failure.}
\label{fig:ratio_large_scale_node_failure}
    \end{minipage}
     \vspace{- 3mm}
\end{figure*}

To evaluate the inter-satellite networking performance under normal conditions, we vary the number of active cells to 10, 20, and 30, and assess both the number of critical satellite nodes and the average end-to-end delay. For each scenario, we set $I_{ij} = 1, \forall \mathbb{C}_i, \mathbb{C}_j \in \mathcal{C}_{\text{active}}$. Specifically, when the number of active ground cells is 10, 20, and 30, there are 45, 190, and 435 active ground cell pairs requiring end-to-end wireless communication services, respectively. 
In all scenarios, we observe that the service connectivity ratios consistently reach $100\%$ in each time window, confirming that the simulated satellite network can fully support the required end-to-end wireless communication services. Figure \ref{fig:num_of_critical_satellite} depicts the average number of critical satellites involved over time, showing that the number of critical satellites fluctuates across different time windows. As the number of ground active cells increases, more critical satellites are required to support these services. Moreover, Figure \ref{fig:num_of_critical_satellite} highlights that larger numbers of active cells result in greater fluctuations in the number of critical satellites over time.
Figure \ref{fig:path_delay_over_time} illustrates the average end-to-end path delay over time under normal conditions. The results show that fluctuations in path delay across different time windows generally remain within the range of 1–2 ms. As the number of services increases, the average end-to-end path delay rises slightly. This is because the addition of active ground cells increases the average distance between cell pairs, leading to longer end-to-end propagation delays.
Since the end-to-end propagation delay of routing paths depends on the distance between cell pairs, the spatial distribution of active cells also affects the average delay. For example, 30 closely located cells may yield lower delays than 10 widely dispersed ones. In our simulation, the 30-cell case includes the 10-cell configuration, and the additional cells are more widely distributed, resulting in a higher average delay.Since our system model analyzes how satellite node importance affects networking performance with a focus on propagation delay, we consider only the widely adopted shortest-path routing to reflect the best networking capability of satellite systems. Additionally, in-orbit packet caching does not affect propagation delay, as the delay depends solely on geographical distance. Although caching may influence transmission delay in applications with specific data sizes, such scenarios are beyond the scope of this work and will be explored in our future studies.

It can be observed from Figure \ref{fig:path_delay_over_time}  that both the red and blue curves (corresponding to 20 and 30 active cells, respectively) exhibit mild rising trends and fluctuations as the time window ID increases, particularly in the first half. These fluctuations gradually diminish in later time windows and remain within 2 ms throughout. This is primarily due to the periodic changes in satellite topology, where longer end-to-end paths may temporarily occur in certain time windows.
It is worth noting that in our simulation, each active ground cell is placed at the geographic center of its corresponding city, with latitude and longitude coordinates extracted from city maps using the STK. Since the importance of satellite nodes and the performance of the satellite network are both time- and service-dependent, different spatial distributions and varying numbers of active ground cell pairs may lead to fluctuations in the simulation results. Nevertheless, the overall performance trends remain consistent across different settings.

Fixing the number of active ground cells at $|\mathcal{C}_{\text{active}}|=30$, we set $I_{ij} = 1, \forall \mathbb{C}_i, \mathbb{C}_j \in \mathcal{C}_{\text{active}}$ to evaluate how the SATB values of critical satellites change over time.
Figure \ref{fig:SATB_over_time} presents the SATB values of the top-8 most critical satellites across time windows. These satellites are identified and ranked based on their SATB values in the first time window. As shown in Figure \ref{fig:SATB_over_time}, the importance of critical satellites fluctuates significantly across different time windows. It can be observed that \textit{satellite\_823} is the most critical (i.e., ranked top-1) node in the first time window but becomes non-critical after \textit{time window\_14}. In contrast, \textit{satellite\_823} is relatively less important than \textit{satellite\_131} in the first time window but experiences a sharp increase in its SATB value in \textit{time window\_18}, indicating that it becomes crucial for supporting an increased number of active cell pairs during \textit{time window\_18}. 
This highlights the dynamic nature of  mega-satellite constellations, where the importance of individual satellites evolve over time, and no single satellite constantly maintains a dominant role.

\subsubsection{Evaluation of impact on routing under node failure event}
Fixing the number of active ground cells at $|\mathcal{C}_{\text{active}}| = 30$, we introduce the node failure event in the first time window to assess how such events impact inter-satellite networking performance. Figure \ref{fig:ratio_single_node_failure} shows the variation in service connectivity ratios under the occurrence of single node failures. We again consider the top-8 most critical satellites in \textit{time window\_1}, with each curve corresponding to the result of a single satellite's failure. Recall that, without node failures, the service connectivity ratios in each time window are $100\%$. 
As seen in Figure \ref{fig:ratio_single_node_failure}, single node failures generally cause short-term service disruptions, indicated by a drop in service connectivity ratios, particularly in the first and subsequent time windows. However, the connectivity ratios gradually recover to $100\%$ as the network topology evolves, highlighting the inherent resilience of the satellite network's dynamic topology against node failures. Notably, the failure of \textit{satellite\_131} causes a significant drop in the service connectivity ratio in \textit{time window\_18}. This behavior is expected and consistent with Figure \ref{fig:SATB_over_time}, where \textit{satellite\_131} has the SATB value ranked eighth in the first time window but experiences a spike during the \textit{time window\_18}. These results confirm that the proposed SATB metric can accurately reflect satellite node criticality and its impact on service connectivity.

Next, we introduce multiple node failure events in the first time window and present the resulting service connectivity ratios across different time windows in Figure \ref{fig:ratio_multiple_node_failure}. In Figure \ref{fig:ratio_multiple_node_failure}, ``Top-k Failure'' refers to the simultaneous failure of the top-k most critical satellites. As expected, the trends in Figure \ref{fig:ratio_multiple_node_failure} are similar to those observed in Figure \ref{fig:ratio_single_node_failure}. When multiple nodes fail, the service connectivity ratio drops sharply, with the severity of the drop increasing as the number of failed nodes rises. However, as time progresses, the service connectivity ratio gradually recovers to $100\%$. This highlights the satellite constellation's strong resilience under multiple node failures.

By introducing larger-scale node failures, Figure \ref{fig:ratio_large_scale_node_failure} illustrates the service connectivity ratios across different time windows under large-scale node failure events, considering scenarios where the top-10 to top-100 most critical satellites fail. Compared to Figure \ref{fig:ratio_multiple_node_failure}, Figure \ref{fig:ratio_large_scale_node_failure} reveals a more pronounced drop in service connectivity ratios as the number of failed nodes increases. Nevertheless, the connectivity ratios  are still able to gradually recover over time, exceeding $95\%$ after approximately $25$ time windows. This underscores the satellite constellations can autonomously restore most of its service connectivity even under large-scale node failures.

\begin{figure}
	\centering
    	\includegraphics[width=60mm]{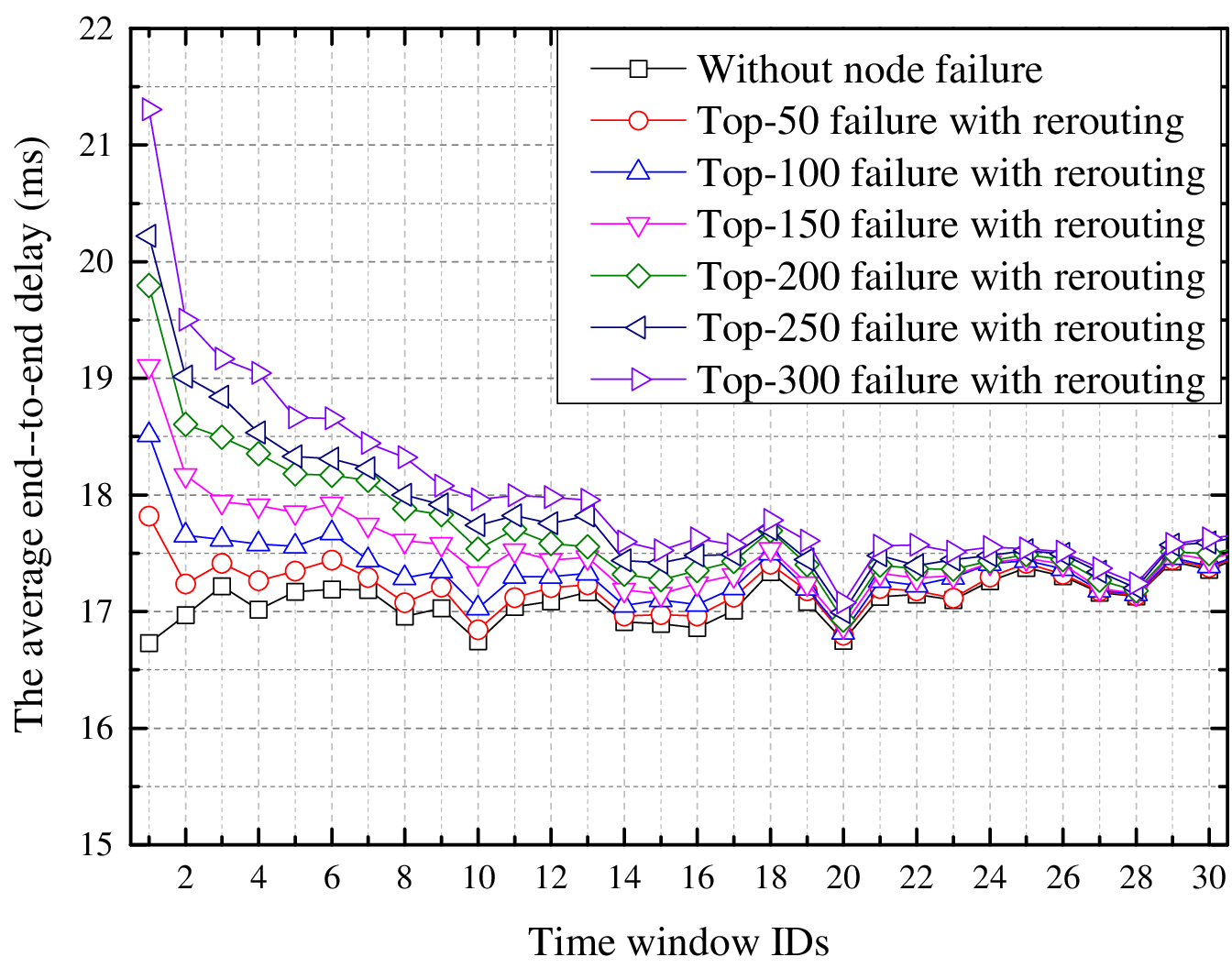}\\
%    	 \vspace{-3 mm}
	\caption{The average end-to-end path delays  under large-scale node failure with rerouting mechanism enabled.}
\label{fig:delay_large_scale_node_failure}
\end{figure}

To further investigate the impact of node failures on satellite constellations with agile rerouting mechanism, we simulate large-scale node failure scenarios involving the top-50 to top-300 most critical nodes while performing rerouting after failures. 
Specifically, when node failures occur, the rerouting mechanism recalculates paths to bypass all the  failed nodes. Remarkably, under all simulated cases, even with the failure of 300 satellites, the service connectivity ratio can be fully restored to $100\%$ with the help of rerouting. To evaluate the impact of node failures on service delay after rerouting, Figure \ref{fig:delay_large_scale_node_failure} presents the average end-to-end path delay after rerouting. It can be observed that a higher number of failed satellites results in increased average path delay during the initial time windows. However, after around 20 time windows, the average path delay gradually returns to pre-failure levels. 
This indicates that the rerouting mechanism, combined with the network's inherent resilience, plays a crucial role in effectively mitigating the impact of large-scale node failures, ensuring rapid connectivity restoration and performance recovery.

\begin{figure}
	\centering
	\includegraphics[width=60mm]{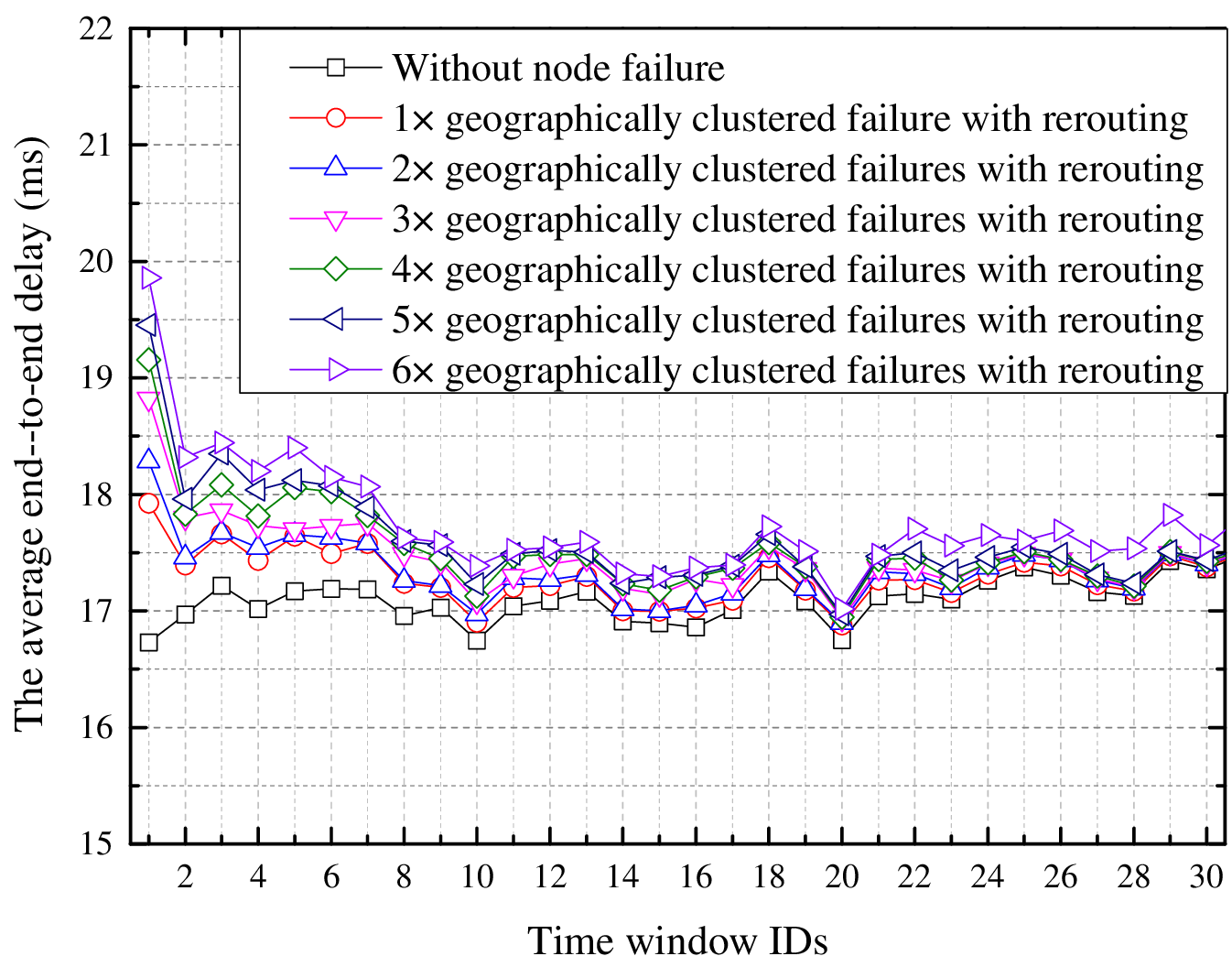}\\
	\caption{The average end-to-end path delays under geographically clustered node failure event with rerouting mechanism enabled.}
\label{fig:delay_geo_failure}
% \vspace{-3 mm}
\end{figure}

\subsubsection{Evaluation of impact on routing under geographically clustered node failure event}
To explore real-world node failure scenarios caused by natural disasters such as solar wind, we simulate geographically clustered node failures involving one to six affected regions, with rerouting performed after each failure event. Each failure region is randomly centered above a ground cell at an altitude of 1100 km, with a spherical radius of 2000 km \cite{ouyang2023mega}. Satellites within the region are considered failed, and the corresponding communication links are removed. The rerouting mechanism then updates paths to bypass the failed nodes. We evaluate the average end-to-end path delay after rerouting under geographically clustered node failures, as shown in Figure~\ref{fig:delay_geo_failure}. In the legend, ``$1 \times$'' and ```$2 \times$'' indicate one and two geographically clustered failure regions in the topology, forming one and two holes in the satellite constellation, respectively. 
To ensure statistical reliability, 95\% confidence intervals are computed for the average delay in each time window. Although omitted from the figure to  avoid visual clutter caused by overlapping curves, the margins of error remain consistently within $\pm$2.98\% of the mean, confirming the low variability and statistical robustness of the results.
The figure reveals a trend similar to Figure~\ref{fig:delay_large_scale_node_failure}, where a higher number of failures leads to increased delay. However, the average delay returns to pre-failure levels after about 10 time windows, which is faster than in the large-scale node failure case. This indicates that geographically clustered node failures cause only localized and short-term disruptions. As time progresses, failed satellites are replaced by others passing over the region, thereby quickly restoring connectivity between ground cells.

\vspace{- 3mm}
\section{Conclusion}  \label{sec:conclusion}

Given the high risks of satellite node failures, this paper targeted on how do these failures impact inter-satellite networking performance. To explore this, we have represented the mega-satellite constellation as the temporal graph and model node failure events accordingly. We have proposed the SATB metric to quantify time-varying node importance for targeted wireless communication services. Leveraging the SATB, we have proposed the analytical framework to identify critical satellites and assess the evolving impact of node failures on inter-satellite networking. Simulations on the Starlink system have revealed two insights for satellite system design. On the one hand, we have found that the dynamic topology of satellite networks inherently helps restore partial service connectivity and mitigate the long-term impact of node failures. On the other hand, we have demonstrated that rerouting mechanisms are essential for unlocking the full resilience potential of satellite constellations to address large-scale node failures and achieve rapid recovery of networking performance.
Future work will explore how alternative constellation topologies and link-type-aware rerouting strategies that distinguish between intra-plane, inter-plane, and cross-plane links may affect network performance and fault tolerance beyond shortest-path routing. Additionally, we aim to incorporate energy consumption and resource utilization models to provide a more comprehensive evaluation of recovery costs under various real-world failure scenarios.

\bibliographystyle{IEEEtran}
\bibliography{reference}

\begin{thebibliography}{10}
\providecommand{\url}[1]{#1}
\csname url@samestyle\endcsname
\providecommand{\newblock}{\relax}
\providecommand{\bibinfo}[2]{#2}
\providecommand{\BIBentrySTDinterwordspacing}{\spaceskip=0pt\relax}
\providecommand{\BIBentryALTinterwordstretchfactor}{4}
\providecommand{\BIBentryALTinterwordspacing}{\spaceskip=\fontdimen2\font plus
\BIBentryALTinterwordstretchfactor\fontdimen3\font minus
  \fontdimen4\font\relax}
\providecommand{\BIBforeignlanguage}[2]{{%
\expandafter\ifx\csname l@#1\endcsname\relax
\typeout{** WARNING: IEEEtran.bst: No hyphenation pattern has been}%
\typeout{** loaded for the language `#1'. Using the pattern for}%
\typeout{** the default language instead.}%
\else
\language=\csname l@#1\endcsname
\fi
#2}}
\providecommand{\BIBdecl}{\relax}
\BIBdecl

\bibitem{le2024survey}
T.~T.~T. Le, N.~U. Hassan, X.~Chen, M.-S. Alouini, Z.~Han, and C.~Yuen, ``A
  survey on random access protocols in direct-access leo satellite-based iot
  communication,'' \emph{IEEE Commun. Surv. Tutorials}, Apr. 2024.

\bibitem{di2019ultra}
B.~Di, L.~Song, Y.~Li, and H.~V. Poor, ``Ultra-dense {LEO}: Integration of
  satellite access networks into {5G} and beyond,'' \emph{IEEE Wireless
  Commun.}, vol.~26, no.~2, pp. 62--69, Apr. 2019.

\bibitem{al2022next}
B.~Al~Homssi, A.~Al-Hourani, K.~Wang, P.~Conder, S.~Kandeepan, J.~Choi,
  B.~Allen, and B.~Moores, ``Next generation mega satellite networks for access
  equality: Opportunities, challenges, and performance,'' \emph{IEEE Commun.
  Mag.}, vol.~60, no.~4, pp. 18--24, Apr. 2022.

\bibitem{hassan2020dense}
N.~U. Hassan, C.~Huang, C.~Yuen, A.~Ahmad, and Y.~Zhang, ``Dense small
  satellite networks for modern terrestrial communication systems: Benefits,
  infrastructure, and technologies,'' \emph{IEEE Wireless Commun.}, vol.~27,
  no.~5, pp. 96--103, Oct. 2020.

\bibitem{SPACEFLIGHT-NOW2022}
SPACEFLIGHT-NOW, ``Spacex passes 2,500 satellites launched for starlink
  internet network,''
  \url{https://spaceflightnow.com/2022/05/13/spacex-passes-2500-satellites-launched-for-companys-starlink-network/},
  2022.

\bibitem{zhang2023time}
N.~Zhang, Z.~Na, J.~Tao, B.~Lin, N.~Zhang, and K.~Zhao, ``Time-varying graph
  and binary tree search based routing algorithm for {LEO} satellite
  networks,'' \emph{IEEE Trans. Veh. Tech.}, vol.~72, no.~10, pp.
  13\,683--13\,688, Oct. 2023.

\bibitem{handley2019using}
M.~Handley, ``Using ground relays for low-latency wide-area routing in
  megaconstellations,'' in \emph{Proc. 18th ACM Workshop on Hot Topics in
  Networks}, New York, NY, Nov. 2019, pp. 125--132.

\bibitem{guo2024time}
Y.~Hu, B.~Guo, C.~Yang, and Z.~Han, ``Time-deterministic networking for
  satellite-based internet-of-things services: Architecture, key technologies,
  and future directions,'' \emph{IEEE Netw.}, vol.~38, no.~4, pp. 111--118,
  Mar. 2024.

\bibitem{zech2015lct}
H.~Zech, F.~Heine, D.~Tr{\"o}ndle, S.~Seel, M.~Motzigemba, R.~Meyer, and
  S.~Philipp-May, ``{LCT for EDRS}: {LEO to GEO} optical communications at 1, 8
  {Gbps} between {Alphasat} and {Sentinel} 1a,'' in \emph{Unmanned/Unattended
  Sensors and Sensor Networks XI; and Advanced Free-Space Optical Communication
  Techniques and Applications}, vol. 9647.\hskip 1em plus 0.5em minus
  0.4em\relax SPIE, Oct. 2015, pp. 85--92.

\bibitem{guo2023online}
B.~Guo, H.~Li, Z.~Zhang, and Y.~Yan, ``Online network slicing for real time
  applications in large-scale satellite networks,'' in \emph{IEEE Int. Conf.
  Commun.}, Rome, Italy, Jun. 2023, pp. 5564--5569.

\bibitem{cao2023edge}
X.~Cao, B.~Yang, Y.~Shen, C.~Yuen, Y.~Zhang, Z.~Han, H.~V. Poor, and L.~Hanzo,
  ``Edge-assisted multi-layer offloading optimization of {LEO}
  satellite-terrestrial integrated networks,'' \emph{IEEE J. Sel. Areas
  Commun.}, vol.~41, no.~2, pp. 381--398, Feb. 2023.

\bibitem{guo2024lightweight}
B.~Guo, Z.~Xiong, Z.~Han, C.~Yuen, and S.~Sun, ``Lightweight maximum-capacity
  path selection for delay-sensitive applications in large-scale leo satellite
  networks,'' \emph{IEEE Trans. Veh. Tech.}, pp. 1--13, Dec. 2024.

\bibitem{lai2022spacertc}
Z.~Lai, W.~Liu, Q.~Wu, H.~Li, J.~Xu, and J.~Wu, ``Space{RTC}: Unleashing the
  low-latency potential of mega-constellations for real-time communications,''
  in \emph{Proc. IEEE Conf. on Comput. Commun. (INFOCOM)}, London, UK, Jun.
  2022, pp. 1339--1348.

\bibitem{guo2024enhanced}
B.~Guo, Z.~Chang, Z.~Han, and Z.~Xiong, ``Enhanced time discretization for
  temporal graph-based continuous service provisioning in large-scale satellite
  networks,'' \emph{IEEE Wireless Commun. Lett.}, vol.~13, no.~9, pp.
  2625--2629, Sep. 2024.

\bibitem{dakic2023delay}
K.~Dakic, C.~C. Chan, B.~Al~Homssi, K.~Sithamparanathan, and A.~Al-Hourani,
  ``On delay performance in mega satellite networks with inter-satellite
  links,'' in \emph{IEEE Global Commun. Conf.}, Kuala Lumpur, Malaysia, Dec.
  2023, pp. 4896--4901.

\bibitem{huang2024fair}
C.~Huang, G.~Chen, P.~Xiao, J.~A. Chambers, and W.~Huang, ``Fair resource
  allocation for hierarchical federated edge learning in space-air-ground
  integrated networks via deep reinforcement learning with hybrid control,''
  \emph{IEEE J. Sel. Areas Commun.}, vol.~42, no.~12, pp. 3618--3631, Dec.
  2024.

\bibitem{zhang2023modeling}
L.~Zhang, Y.~Du, and Z.~Sun, ``Modeling and analysis of cascading failures in
  leo satellite networks,'' \emph{IEEE Trans. Netw. Science and Engineering},
  vol.~11, no.~1, pp. 807--822, Feb, 2023.

\bibitem{yue2023low}
P.~Yue, J.~An, J.~Zhang, J.~Ye, G.~Pan, S.~Wang, P.~Xiao, and L.~Hanzo, ``Low
  earth orbit satellite security and reliability: Issues, solutions, and the
  road ahead,'' \emph{IEEE Commun. Surveys \& Tutorials}, vol.~25, no.~3, pp.
  1604--1652, Aug. 2023.

\bibitem{yue2020outage}
X.~Yue, Y.~Liu, Y.~Yao, T.~Li, X.~Li, R.~Liu, and A.~Nallanathan, ``Outage
  behaviors of noma-based satellite network over shadowed-rician fading
  channels,'' \emph{IEEE Trans. Veh. Tech.}, vol.~69, no.~6, pp. 6818--6821,
  Apr. 2020.

\bibitem{li2024performance}
K.~Li, J.~Wang, T.~Hou, A.~Li, X.~Yue, Y.~Liu, and W.~Chen, ``Performance
  analysis of oma/noma-aided satellite communication networks: A stochastic
  geometry approach,'' \emph{IEEE Trans. Commun.}, Dec. 2024.

\bibitem{tao2023joint}
J.~Tao, Z.~Na, B.~Lin, and N.~Zhang, ``A joint minimum hop and earliest arrival
  routing algorithm for leo satellite networks,'' \emph{IEEE Trans. Veh.
  Tech.}, vol.~72, no.~12, pp. 16\,382--16\,394, Dec. 2023.

\bibitem{chen2022leo}
Q.~Chen, L.~Yang, D.~Guo, B.~Ren, J.~Guo, and X.~Chen, ``{LEO} satellite
  networks: When do all shortest distance paths belong to minimum hop path
  set?'' \emph{IEEE Trans. Aerosp. Electron. Syst.}, vol.~58, no.~4, pp.
  3730--3734, Aug. 2022.

\bibitem{chen2024shortest}
Q.~Chen, L.~Yang, Y.~Zhao, Y.~Wang, H.~Zhou, and X.~Chen, ``Shortest path in
  {LEO} satellite constellation networks: An explicit analytic approach,''
  \emph{IEEE J. Sel. Areas Commun.}, vol.~42, no.~5, pp. 1175--1187, May 2024.

\bibitem{handley2018delay}
M.~Handley, ``Delay is not an option: Low latency routing in space,'' in
  \emph{Proc. 17th ACM Workshop on Hot Topics Netw.}, New York, NY, Nov. 2018,
  pp. 85--91.

\bibitem{chen2021analysis}
Q.~Chen, G.~Giambene, L.~Yang, C.~Fan, and X.~Chen, ``Analysis of
  inter-satellite link paths for {LEO} mega-constellation networks,''
  \emph{IEEE Trans. Veh. Tech.}, vol.~70, no.~3, pp. 2743--2755, Mar. 2021.

\bibitem{wang2019pkn}
S.~Wang, Y.~Zhao, and H.~Xie, ``Pkn: Improving survivability of leo satellite
  network through protecting key nodes,'' in \emph{Proc. 15th Int. Conf.
  emerging Netw. EXperiments Tech.}, Orlando, FL, Dec. 2019, pp. 7--8.

\bibitem{han2021secure}
R.~Han, L.~Bai, C.~Jiang, J.~Liu, and J.~Choi, ``A secure architecture of
  relay-aided space information networks,'' \emph{IEEE Netw.}, vol.~35, no.~4,
  pp. 88--94, Aug. 2021.

\bibitem{li2022secure}
H.~Li, D.~Shi, W.~Wang, D.~Liao, T.~R. Gadekallu, and K.~Yu, ``Secure routing
  for {LEO} satellite network survivability,'' \emph{Computer Netw.}, vol. 211,
  no.~C, pp. 1389--1286, Jul. 2022.

\bibitem{lai2023achieving}
Z.~Lai, H.~Li, Y.~Wang, Q.~Wu, Y.~Deng, J.~Liu, Y.~Li, and J.~Wu, ``Achieving
  resilient and performance-guaranteed routing in space-terrestrial integrated
  networks,'' in \emph{IEEE Conf. Computer Commun. ( INFOCOM)}.\hskip 1em plus
  0.5em minus 0.4em\relax New York, NY: IEEE, May 2023, pp. 1--10.

\bibitem{zhang2024resilient}
Y.~Zhang, K.~Zhao, and W.~Li, ``A resilient routing algorithm for handling $ k
  $-link/node failure in inclined {LEO} mega-constellations,'' \emph{IEEE
  Trans. Aerosp. Electron. Syst.}, vol.~60, no.~4, pp. 4876--4886, Mar. 2024.

\bibitem{boccaletti2006complex}
S.~Boccaletti, V.~Latora, Y.~Moreno, M.~Chavez, and D.-U. Hwang, ``Complex
  networks: Structure and dynamics,'' \emph{Physics Reports}, vol. 424, no.
  4-5, pp. 175--308, Feb. 2006.

\bibitem{costa2007characterization}
L.~d.~F. Costa, F.~A. Rodrigues, G.~Travieso, and P.~R. Villas~Boas,
  ``Characterization of complex networks: A survey of measurements,''
  \emph{Advances in physics}, vol.~56, no.~1, pp. 167--242, Apr. 2007.

\bibitem{lu2020structural}
Y.~Lu, G.~Min, Z.~Zuo, R.~Liang, and Z.~Duan, ``Structural performance of
  satellite networks: a complex network perspective,'' \emph{IEEE Systems J.},
  vol.~15, no.~3, pp. 3848--3859, Jul. 2020.

\bibitem{zhao2020efficient}
C.~Zhao, J.~Liu, M.~Sheng, and Y.~Dai, ``Efficient betweenness based content
  caching and delivery strategy in wireless networks,'' \emph{IEEE Wireless
  Commun. Letters}, vol.~9, no.~9, pp. 1547--1551, Sep. 2020.

\bibitem{xu2023robustness}
X.~Xu, Z.~Gao, and A.~Liu, ``Robustness of satellite constellation networks,''
  \emph{Computer Commun.}, vol. 210, pp. 130--137, Oct. 2023.

\bibitem{zhang2017temporal}
Z.~Zhang, C.~Jiang, S.~Guo, Y.~Qian, and Y.~Ren, ``Temporal centrality-balanced
  traffic management for space satellite networks,'' \emph{IEEE Trans. Veh.
  Tech.}, vol.~67, no.~5, pp. 4427--4439, May 2017.

\bibitem{bakhsh2024multi}
Z.~M. Bakhsh, Y.~Omid, G.~Chen, F.~Kayhan, Y.~Ma, and R.~Tafazolli,
  ``Multi-satellite {MIMO} systems for direct satellite-to-device
  communications: A survey,'' \emph{IEEE Commun. Surv. Tutor.}, pp. 1--29, Aug.
  2024.

\bibitem{araniti2015contact}
G.~Araniti, N.~Bezirgiannidis, E.~Birrane, I.~Bisio, S.~Burleigh, C.~Caini,
  M.~Feldmann, M.~Marchese, J.~Segui, and K.~Suzuki, ``Contact graph routing in
  {DTN} space networks: overview, enhancements and performance,'' \emph{IEEE
  Commun. Mag.}, vol.~53, no.~3, pp. 38--46, Mar. 2015.

\bibitem{evans1998satellite}
J.~V. Evans, ``Satellite systems for personal communications,'' \emph{Proc. the
  IEEE}, vol.~86, no.~7, pp. 1325--1341, Aug. 1998.

\bibitem{yao2025leo}
Z.~Yao, J.~Zhou, L.~Xiao, M.~Feng, P.~Xiao, and T.~Jiang, ``{LEO}
  intersatellite communications: From routing to resource allocation,''
  \emph{IEEE Vehicular Technology Magazine}, pp. 1--12, Jan. 2025.

\bibitem{ouyang2023mega}
Q.~Ouyang, N.~Ye, S.~Miao, B.~Kang, A.~Wang, and L.~Zhao, ``Mega constellation
  networks are reliable against geographical failure,'' in \emph{IEEE Vehicular
  Technology Conf. (VTC2023-Fall)}.\hskip 1em plus 0.5em minus 0.4em\relax Hong
  Kong, China: IEEE, Dec. 2023, pp. 1--5.

\end{thebibliography}

%\begin{thebibliography}{1}

%0.5em minus 0.4em\relax Harlow, England: Addison-Wesley, 1999.

%\end{thebibliography}


% if you will have a photo at all:
%\begin{IEEEbiography}{John Doe}
%	Biography text here.
%\end{IEEEbiography}


%\newpage



%\vfill


\end{document}